\documentstyle[prc,preprint,aps,epsfig]{revtex}
\begin{document}
 \input FEYNMAN

\baselineskip=1. \baselineskip
\draft
\preprint{YITP-97-13}
\title{Particle production in quantum transport theories}

\author{     P. Bo\.{z}ek~\footnote[1]{electronic
mail~:~bozek@solaris.ifj.edu.pl}  }
\address{ Yukawa Institute for Theoretical Physics, Kyoto University,
Kyoto 606-01 , Japan \\  and \\
 Institute of Nuclear Physics, 31-342 Krak\'{o}w, Poland}
\date{\today}
\maketitle 

\begin{abstract}
The particle production in the intermediate energy heavy ion
collisions is discussed in the framework of the nonequilibrium
Green's functions formalism. The evolution equations of the Green's functions
for fermions allows for the discussion of the off-shell fermion propagator
and of the large momentum component in the initial state.
For the case of a homogeneous system  numerical calculations
of the meson production rate  are
performed and  compared with the 
semiclassical production rate.
\end{abstract}

\pacs{24.10-i, 25.75Dw}

\narrowtext

\section{Introduction}

The production of particles in the interacting nuclear medium has been 
a rich field of study \cite{niemcy}. The
particle production is a quantity which can be to a large extent
estimated in theoretical models. The particle production was  used
as a tool of the study of  the highly nonequilibrium dynamics in
the early stage of the nuclear collision and also as a probe of the
behavior and a characteristic of the almost equilibrated system reached
in the course of the collision. The estimation of the particle production
probability usually is based on the free production cross section
with some possible medium modifications included. This production
cross section, taken for the 2-nucleon collision process, is used in the
semiclassical Boltzmann-Uehling-Uhlenbeck (BUU) 
type simulations of the nuclear dynamics. 
A completely different approach to the particle production emerges 
from the quantum transport equations. The importance of the quantum 
transport description was realized in the correct treatment of the
Landau-Pomeranchuk-Migdal effect in medium \cite{kn_vo}. 
The approach based on the quantum transport equations takes into
account the finite time extent between successive collisions and
regularizes the  divergent cross section for long-wavelength photons
which require large formation time.

Two problems related to the
 particle production in intermediate energy heavy ion collisions
are particularly interesting~: the subthreshold meson production and the 
medium modifications of the meson properties. The production of the 
particles below the nucleon-nucleon energy threshold requires the
presence of  some
additional mechanism in nuclear medium allowing for the production.
 Besides the obvious Fermi
motion of nucleons,
additional mechanisms could include the initial high momentum component 
in the nucleus and the many nucleon collisions. These phenomena can
be treated in a unified way in the quantum nonequilibrium Green's
function formalism \cite{kb,d1,ma}.
 The medium modification of the meson properties
include the possible mass shift and width modifications in medium. 
If this occurs,
a natural question arises also about possible strong medium modifications
of the production rates. In the case of very strong medium
modification
of the meson properties
 the meson  production should be treated in a
quantum transport  framework.

In this work it will appear that the difference between the quantum
and semiclassical production rates is  very important.
Usually one estimates the off-shell effects by taking a generalized
parameterization of the fermion spectral functions in the calculation 
of the meson production rate in 2-nucleon collision.
However, in the quantum transport equation the perturbative expansion 
using resumed fermion propagators is organized differently.
The first nonzero contribution to the meson production comes from the 
one-loop diagram and its contribution is generally large.
Another way of discussion of the subthreshold particle production is
to estimate the particle production in
  3 or more nucleon collisions processes. 
A large class of such meson production processes involving 
many nucleons is included
in the one-loop meson self-energy with interacting off-shell fermion
propagators.

In the present work in order to make a detailed estimation of the
quantum
and semiclassical production rates in the same system we will use a 
simple model.
In this model of self-interacting nonrelativistic fermions weakly
coupled to mesons we will calculate the one-loop quantum meson production
rate and the corresponding semiclassical rate from the 2-nucleon
collision process. Clearly we cannot use the free on shell meson
production cross section as a starting point. Instead we will use 
a  simple meson-nucleon 
coupling and we will derive the corresponding semiclassical 
production rate.
The numerical calculations of the dynamics of the
fermion Green's functions  are done in a similar way as in the work of 
Danielewicz \cite{d2}. The production rate of mesons is estimated in a 
perturbative way, i.e. we assume that it is small and that the meson
production does not modify the fermion dynamics.

In order to make the calculation feasible we  assume that the system
is spatially homogeneous. The quantum evolution equations are much
simpler in this case, since the momentum integrals are the same as in the 
usual  evaluation of perturbative  diagrams 
and only the time integration must be done on a
contour in the complex time plane.
 From the experimental evidence it seems as well established 
 that the particle
production around the threshold, but sufficiently above the absolute
threshold, is an incoherent process. The number of produced mesons
scales very well with the number of participants. The result of the
calculation in
the homogeneous system giving the particle production rate per unit
volume can  be scaled to the  size of the actual system under
consideration.
Thus, we believe  that our results could be used for actual estimates of
the production probability. Of course this would require the use of
 realistic meson nucleon vertices, and also  the inclusion of the
delta degrees of freedom.
Our goal in this work is restricted to the presentation of the 
particle production in the nonequilibrium transport equations, with
comparison to
the semiclassical production rates. However, as a good motivation
one should  keep in mind the fact that the
comparison of the particle production estimated in the off-shell Green's
functions framework would represent 
a powerful test of the quantum effects.
The particle production is one of the rare observables which can be
reasonably well estimated in a homogeneous system and then compared to the
experimental production rate per participant. Only the estimation of the
absorption of the  produced mesons  must be done in a finite system. 
The difference in the transport coefficients  between the quantum and the 
semiclassical
evolutions discussed up to now \cite{d2,ma2,koh,he} is
not directly observable and could be effectively included in
a medium modification 
of the nucleon-nucleon cross section.

\section{Particle production in Quantum transport equations}

We will consider a system of nonrelativistic interacting   fermions.
The first numerical solution of the Kadanoff-Baym equations
for a homogeneous system was  presented in Ref.  \cite{d2}.
For the dynamics of the fermions we follow the same approach.
In sections \ref{mp} and \ref{nmp} 
we will calculate the meson production rate in the system of interacting
off-shell fermions.

\subsection{Quantum transport equations for fermions}
\label{qte}

\begin{figure}
  \begin{picture}(10000,18000)
  \thicklines
  \put(200,9000){\vector(1,0){21000}}
   \drawline\fermion[\S\REG](1200,13000)[3800]
   \drawline\fermion[\E\REG](1200,9200)[17000]
   \drawline\fermion[\E\REG](1200,8800)[17000]
   \drawline\fermion[\S\REG](1200,8800)[3800]
   \drawline\fermion[\S\REG](18200,9200)[400]
  \put(1000,13000){\makebox(1000,1000)\Huge{$t_0+i \tau_0$}}
  \put(1000,5000){\makebox(1000,1000)\Huge{$t_0-i \tau_0$}}
  \end{picture}
\caption{
The contour in the complex time plane on which the Green's functions
are defined.}
\label{contour}
\end{figure}
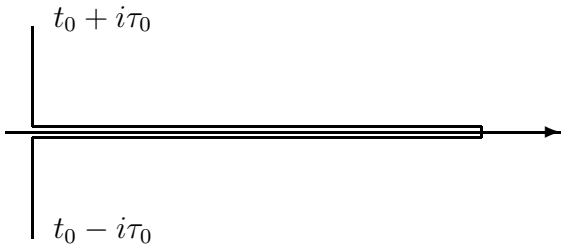

In this section we recall the notation of \cite{kb,d1} and the numerical 
procedure for the solution of the nonequilibrium dynamics of 
fermion Green's functions. 
The Hamiltonian of fermions interacting with a two-body potential $V$,
 can be written using the annihilation and creation operators 
$\hat{\psi}(x)$ and $\hat{\psi}^{\dag}(x) $ as follows~:
\begin{equation}
\label{ham}
\hat{H}= \int dx \ \hat{\psi}^{\dag}(x) \frac{-\nabla^{2}}{2 m}\hat{\psi}(x)
+\frac{1}{2}\int dx \int dy \ \hat{\psi}^{\dag}(x)\hat{\psi}^{\dag}(y)
V(x-y)\hat{\psi}(y)\hat{\psi}(x) \ .
\end{equation}
The spin and isospin indices are not written explicitly and we take a
scalar potential for simplicity.
The one-body evolution of the system can be described by the Green's
functions~:
\begin{eqnarray}
\label{gre_fun}
G^{<}(x_1,t_1,x_2,t_2) & = &
i<\hat{\psi}^{\dag}_H(x_2,t_2)\hat{\psi}_H(x_1,t_1)>
\ , \nonumber \\
G^{>}(x_1,t_1,x_2,t_2) & = &
-i<\hat{\psi}_H(x_1,t_1)\hat{\psi}^{\dag}_H(x_2,t_2)> \ ,
\end{eqnarray}
where the field operators $\psi_H$ are in the Heisenberg representation
and the average $< \dots>$ is the average over the initial state.
These Green's functions are a particular case of  the general
contour time ordered Green's function~:
\begin{equation}
G(x_1,t_1,x_2,t_2) = i < T_C
\hat{\psi}^{\dag}_H(x_2,t_2)\hat{\psi}_H(x_1,t_1)>
 \ ,
\end{equation}
where  the field operators are ordered according to the time on the
contour in the complex plain (Fig \ref{contour}).
For $t_1$, $t_2$ on different branches of the real time part of the
contour the Green's function $G$ can be replaced by two Green's
functions
$G^{<(>)}$ defined on the real line.
In a spatial homogeneous system there is no dependence on the 
macroscopic coordinate $\frac{x_1+x_2}{2}$, and after Fourier
transforming in the relative coordinate $x_1-x_2$ we obtain the Green's 
function in momentum space $G^{<(>)}(p,t_1,t_2)$~.
The evolution of the system starting at some time $t_0$ can be
described by the Schwinger-Dyson  equations~:
\begin{eqnarray}
\label{kb}
i\frac{\partial}{\partial t_1} G(p,t_1,t_2)-\omega_p G(p,t_1,t_2)& = &
\delta_C(t_1,t_2) +
\int_C dt^{'} \Sigma(p,t_1,t^{'})G(p,t^{'},t_2) \nonumber \\ 
-i\frac{\partial}{\partial t_2} G(p,t_1,t_2)-\omega_p G(p,t_1,t_2)& = &
\delta_C(t_1,t_2) +
\int_C dt^{'} G(p,t_1,t^{'})\Sigma(p,t^{'},t_2)
\ ,
\end{eqnarray}
where $\Sigma(p,t_1,t_2)$ is the self-energy which must be calculated
in some approximation, $\delta_C(t_1,t_2)$ is the delta function on
the contour and the time integration $t^{'}$ is done on the contour
(Fig. \ref{contour}).
We neglect the mean field potential and the medium modification of the 
nucleon effective mass, so that $\omega_p=\frac{p^2}{2m}$.
For $t_1$, $t_2$ real and on different branches of the contour we obtain
the Kadanoff-Baym equations~:
\begin{eqnarray}
\label{Ka_Ba}
i\frac{\partial}{\partial t_1} G^{<}(p,t_1,t_2)-\omega_p G^{<}(p,t_1,t_2)& = &
\left(\int_{t_0+i\tau_0}^{t_0} + \int_{t_0}^{t_1}\right) dt^{'}
\Sigma^{>}(p,t_1,t^{'})G^{<}(p,t^{'},t_2) \nonumber \\
& & +  \int_{t_1}^{t_2}
dt^{'}\Sigma^{<}(p,t_1,t^{'})G^{<}(p,t^{'},t_2)
\nonumber \\ 
& & + 
\left(\int_{t_0}^{t_0-i\tau_0} + \int_{t_2}^{t_0} \right)dt^{'}
\Sigma^{<}(p,t_1,t^{'})G^{>}(p,t^{'},t_2) \ ,
\nonumber \\ 
i\frac{\partial}{\partial t_1} G^{>}(p,t_1,t_2)-\omega_p G^{>}(p,t_1,t_2)& = &
\left(\int_{t_0+i\tau_0}^{t_0} + \int_{t_0}^{t_2}\right) dt^{'}
\Sigma^{>}(p,t_1,t^{'})G^{<}(p,t^{'},t_2) \nonumber \\
& & +  \int_{t_2}^{t_1}
dt^{'}\Sigma^{>}(p,t_1,t^{'})G^{>}(p,t^{'},t_2)
\nonumber \\ 
& & + 
\left(\int_{t_0}^{t_0-i\tau_0} + \int_{t_1}^{t_0}\right) dt^{'}
\Sigma^{<}(p,t_1,t^{'})G^{>}(p,t^{'},t_2) \ ,
\end{eqnarray}
and similar equations for the evolution in $t_2$. 
The description of the fermion evolution using only the one-body
fermion Green's function involves the assumption that there are no
2 or many nucleon correlations present at the initial time
\cite{hall,d1}. 
  Such many-nucleon correlations could  be
important for some processes involving nucleons in nuclei \cite{fujita}.

The actual solution
requires first the solution  on the imaginary time part of the
contour, which is done iteratively. Then the Kadanoff-Baym equations 
are solved on the real time part of the contour. Also similar
equations are solved for the case where one of the time arguments
 of the Green's function is on the real part 
and the other one is on the
imaginary part of the contour. Of course the Hamiltonian of 
the evolution on the imaginary
part of the contour must be modified. Otherwise we would always 
obtain the ground state of the one-body evolution with the Hamiltonian
(\ref{ham}), for $\tau_0 \rightarrow \infty$.
 This can be done as an additional constraint or some
modification of the Hamiltonian on the imaginary part of the contour.
We will describe the nuclear collision as the equilibration of two 
counter-streaming flows of nuclear matter. Accordingly we choose,
following \cite{d2}, as the initial state
two Fermi spheres of radius $p_f=255MeV/c$ centered at $\pm \frac
{p_{coll}}{2}$ with the kinetic energy of the Hamiltonian on the imaginary 
part of the contour $\frac{p_T^2+(|p_L|-p_{coll}/2)^2}{2 m}$
 instead of $\frac{p^2}{2 m}$ on the real part of the contour.
The resulting evolution of the fermion Green's function is the so
called evolution with correlated initial state.
The self energy is taken in the direct Born approximation~:
\begin{eqnarray}
\Sigma^{<(>)}(p,t_1,t_2)& = & 4 \int \frac{d^3p_1}{(2\pi^3)}
\int \frac{d^3p_2}{(2\pi)^3} G^{>(<)}(p_1,t_2,t_1) \nonumber \\
& &
G^{<(>)}(p_2,t_2,t_1)G^{<(>)}(p+p_1-p_2,t_2,t_1) V(p_2-p)^{2} \ ,
\end{eqnarray}
 which
allows for an efficient numerical evaluation.
The factor $4$ comes from spin isospin trace and the potential $V(p)=
\pi^{3/2} \eta^3 V_0 exp(-\frac{p^2}{4 \eta^2})$ ($V_0=453MeV$~,
$\eta=0.57fm$) 
should be thought of as a local
approximation to the $T$ matrix.

Another possibility is to use only the real part of contour for the
calculation of the evolution of the system. The fermion momentum 
distribution at
$t=t_0$ is taken as two Fermi spheres. The kinetic energy in this case 
is increased in the course of the evolution and the final distribution 
corresponds the equilibrium distribution with higher temperature than
in the case of the evolution  including also  the 
imaginary part of the 
contour. This  real time only  quantum transport evolution 
of the fermion system will be refered to as
the evolution from an uncorrelated or a Hartree initial state.

We have used an axially symmetric system for the evolution
of the Kadanoff-Baym equations, with momentum 
grid of $46MeV/c$ and of the size $|p|<920MeV/c$.
The imaginary time evolution was done for $\tau_0=3fm/c$ and the real
time evolution was done for  $20fm/c$. The momentum integrals
are done by a Fourier-Bessel transformation and the time integrals are 
done in a usual numerical integration for each pair of times
$t_1$, $t_2$ ($t_1>t_2$).

\subsection{Meson production}
\label{mp}

The meson production can be analyzed in a model where in addition to
the Hamiltonian $\hat{H}$ for interacting fermions (\ref{ham}) we have a 
 meson-fermion interaction vertex and of course the meson kinetic
energy term.
 For the meson-fermion interaction we could take the form of the interaction 
as in the pion-nucleon case \cite{oset}~:
\begin{equation}
\label{inteham}
\hat{H}_I = g \int dx \left(\hat{\psi(x)}^{\dag} \sigma^i \partial_i 
\hat{\phi}_j(x) \tau^j \hat{\psi(x)} + 
\hat{\psi(x)}^{\dag} \sigma^i \partial_i 
\hat{\phi}_j(x)^{\dag} \tau^j \hat{\psi(x)} \right) \ ,
\end{equation}
where $\hat{\phi}$ is the pion field operator, $\tau$ and $\sigma$
are the isospin and spin matrices. The interaction strength
$g$ incorporates in general a momentum dependent form factor.
However, in our study the detailed structure of the interaction vertex is not
important. Both in the one-loop diagram and in the semiclassical
production rate the meson self-energy acquires the same coefficient
from the meson-fermion interaction vertex. In the case of the
interaction vertex written  above
(\ref{inteham}),  we
have a common factor~:
\begin{equation}
\lambda^2=4 g^2 q^2 \ ,
\end{equation}
in the momentum space, where $q$ is the momentum of the meson.
 The interaction denoted in the following by $\lambda^2$ 
could  be used also for any kind
of meson-nucleon  vertex, after taking the spin isospin trace.
 In the present study, which is a first 
estimation of the quantum production rate, we will not  be interested
in the detailed predictions. Instead we will calculate the rates
of the meson production
scaled by $1/\lambda^2$ for the comparison of  the quantum transport 
equations and the
semiclassical equations. Also we will treat the meson mass as a free
parameter, performing the calculation for several different masses of
produced particles.

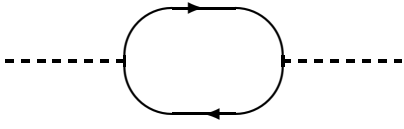
\begin{figure}
  \begin{picture}(10000,18000)
  \thicklines
   \seglength=300 \gaplength=300
   \drawline\scalar[\E\REG](0,10000)[8]
  \addtolength{\scalarbackx}{3000}
  \put(\scalarbackx,\scalarbacky){\oval(6000,4000)} 
  \addtolength{\scalarbacky}{2000}
  \put(\scalarbackx,\scalarbacky){\vector(1,0){0}}
  \addtolength{\scalarbacky}{-4000}
  \put(\scalarbackx,\scalarbacky){\vector(-1,0){0}}
  \addtolength{\scalarbacky}{2000}
  \addtolength{\scalarbackx}{3000}
   \seglength=300 \gaplength=300
     \drawline\scalar[\E\REG](\scalarbackx,\scalarbacky)[8]
  \end{picture}
\caption{The   one-loop meson self-energy diagram. The solid lines are 
the fermion propagators, the dashed lines indicate the location of the 
meson-nucleon vertices.}
\label{one_loo}
\end{figure}

The evolution of the nucleon meson system can be analyzed in the same
way as the system of fermions, discussed in the previous section.
The meson Green's functions can be defined as~:
\begin{eqnarray}
\label{gre_mes}
D^{<}(x_1,t_1,x_2,t_2) & = &
-i<\hat{\phi}^{\dag}_H(x_2,t_2)\hat{\phi}_H(x_1,t_1)>
\ , \nonumber \\
D^{>}(x_1,t_1,x_2,t_2) & = &
-i<\hat{\phi}_H(x_1,t_1)\hat{\phi}^{\dag}_H(x_2,t_2)> \ ,
\end{eqnarray}
where in the following we restrict our study to one meson species.
The meson nucleon interaction induces a backreaction of the produced
mesons on the fermion
evolution. However we will suppose that the meson-nucleon vertex is
weak, and that the meson production does not modify the fermion
dynamics. The fermions do interact by the interaction $V$ which
induces the 
equilibration of fermions and leads to nonzero width of the fermion 
Green's functions. Thus, the  meson production can be calculated from
the
known fermion Green's functions. We can write
also the  Kadanoff-Baym evolution equations for mesons~:
\begin{eqnarray}
\label{kb_mes}
i\frac{\partial}{\partial t_1} D(p,t_1,t_2)-\Omega_p D(p,t_1,t_2)& = &
\delta_C(t_1,t_2) +
\int_C dt^{'} \Pi(p,t_1,t^{'})D(p,t^{'},t_2) \nonumber \\ 
-i\frac{\partial}{\partial t_2} D(p,t_1,t_2)-\Omega_p D(p,t_1,t_2)& = &
\delta_C(t_1,t_2) +
\int_C dt^{'} D(p,t_1,t^{'})\Pi(p,t^{'},t_2)
\ ,
\end{eqnarray}
where $\Pi$ is the meson self-energy and $\Omega_p=\sqrt{m_s^2+p^2}$
is the relativistic meson kinetic energy.
We take at this point the meson kinetic energy in the 
relativistic form in order to have a correct estimate of the energy of 
the meson for any meson mass and momentum. Otherwise the mesons are treated
in an analogous way as the nonrelativistic fermions. The meson
self-energy is taken in the one-loop form (Fig. \ref{one_loo}).
The equations (\ref{kb_mes}) can be used for a complete study of the
meson evolution in the nonequilibrium nuclear matter, including the in 
medium modifications of the meson properties, the meson production
and absorption. 
In the present work we are interested only in the calculation of the meson
production rate. It is equal to  the rate of the change of the equal time
meson Green's function $iD^{<}(p,t,t)={dN(p,t)}/{d^3p}$, which is the
 meson momentum  distribution in a homogeneous system~.
 Supposing that the number of
produced mesons is small, which is true for sufficiently small
$\lambda^2$, we can neglect the meson occupancy on the right hand side 
of the equations (\ref{kb_mes}), taking~:
\begin{eqnarray}
D^{<}_0(p,t_1,t_2) & = & 0 \nonumber \\
D^{>}_0(p,t_1,t_2) & = & - i \exp \left(-i\Omega_p(t_1-t_2) \right) \ .
\end{eqnarray}
This approximation is consistent with the neglect of the meson influence
on the fermion dynamics.
The production rate for mesons takes the form
\begin{eqnarray}
\label{row_mes}
\frac{d N(p,t)}{d^3 p d t} & = & 2 {\cal R}e \Bigg( - \int_{t_0}^{t} dt^{'}
\Pi^{<}(p,t,t^{'}) D^{>}_0(p,t^{'},t) \nonumber \\ & &
+ \int_{t_0}^{t_0-i\tau_0} dt^{'}
\Pi^{<}(p,t,t^{'}) D^{>}_0(p,t^{'},t) \Bigg) \ .
\end{eqnarray}
The meson one-loop self energy is ~:
\begin{equation}
\label{se_ol}
\Pi^{<}(p,t_1,t_2)=  - i \lambda^2 \int \frac{d^3q}{(2\pi)^3}
G^{<}(p-q,t_1,t_2)G^{>}(q,t_2,t_1) \ .
\end{equation}

At this point, some comment is again welcomed concerning the
integration on the 
imaginary part of the contour  in the Kadanoff-Baym equations.
We assume that the imaginary
part of the evolution does not create mesons. This means again that
we modify the Hamiltonian on the imaginary part of the evolution on the 
time contour, or 
equivalently that the evolution operator on the imaginary part of the contour
includes a projection on the states without mesons \cite{lan}.
 In this way we
have no mesons at the initial time $t_0$. The problem of the presence
of mesons in the initial state is by its own very interesting and
could be addressed by the Kadanoff-Baym equations (\ref{kb}) and
(\ref{kb_mes}).
The calculation of the meson distribution in the initial state
 would require a self-consistent solution of the interacting 
meson-fermion system. This is beyond the scope of the present work,
where we restrict ourselves to the perturbative calculation of the
meson production rate from the fermion Green's functions.

The rate of the meson production (\ref{row_mes}) consists of two
time integrals, a real time 
and an imaginary time integration. For $t \rightarrow \infty$ the imaginary 
time integration is negligible, and we obtain as the production rate
the imaginary part of the meson self-energy taken at the energy
$\Omega_q$. For finite integration times, the situation is different.
In particular, taking the stationary fermion Green's functions on-shell~:
\begin{eqnarray}
\label{ons}
G^{<}(p,t_1,t_2) &=&i f(p) \exp \left(-i\omega_p(t_1-t_2) \right) \ ,
\nonumber \\
G^{>}(p,t_1,t_2) &=&-i (1-f(p)) \exp \left(-i\omega_p(t_1-t_2) \right) 
\ ,
\end{eqnarray}
 we obtain from the real part of the time  integration in
Eq. (\ref{row_mes}) an oscillatory behavior~:
\begin{equation}
\frac{dN(q,t)}{d^3qdt} = 2 \lambda^2 \int \frac{d^3p}{(2\pi)^3}   
f(q-p)(1-f(p)) \  \frac{\sin
\left((\Omega_q-\omega_{q-p}+\omega_{p})t\right)}
{\Omega_q-\omega_{q-p}+\omega_{p}} \ ,
\end{equation}
 which only on average gives a
zero production rate.
On the other hand, taking the same form of the on-shell fermion
Green's functions (\ref{ons}) also on the imaginary interval of
the time
integration in Eq. (\ref{row_mes}) and taking $\tau_0 \rightarrow
\infty$ one obtains~:
\begin{equation}
\frac{dN(q,t)}{d^3qdt} = 0 \ ,
\end{equation}
 exactly  for each $t$. Thus, we believe that
the imaginary part of the time integration is important in the
calculation of the production rate in a finite time interval
after $t_0$. Of course in the system that we solve numerically
 we have    off-shell  
Green's function which  gives a nonzero 
production rate. Also in the far from equilibrium system at the
beginning
of the collision the Green's function depend strongly on the
macroscopic time $\frac{t_1+t_2}{2}$~.
One should remember that even in the stationary case the analytical
continuation of the self-energy from the imaginary to the real times
is  nontrivial and in fact it is  possible only in the perturbative
calculation, when the analytical expressions for the self energy are known.
The method presented above is way of calculating the real time
quantities from the  Green's functions evolved
in the imaginary time. It would be interesting to analyze in details 
this problem in the case
of a finite temperature stationary solution, i.e. in the case where
the imaginary evolution is the usual finite temperature evolution over 
an imaginary time interval $1/T$ with the same Hamiltonian on the real and
the imaginary part of the contour.

\subsection{Numerical results for the quantum production rate}
\label{nmp}

The numerical calculation of the fermion evolution was performed 
for the  energy of the collision of $180MeV$ and $110MeV/$nucleon
 in the c.m..
We have calculated two  evolutions of the fermion Green's functions,
corresponding to a correlated or uncorrelated initial state.
The evolution was followed up to $20fm/c$. At that time the 
fermion distribution is almost spherical and equilibrated.
Also, at $t=20fm/c$
there is already no influence of the initial part of the time
contour on the evolution. However, the evolution for 
the initial correlated and the evolution for the initial
Hartree state have different  temperatures at large times. The
temperature in the second case is larger and so is the meson
production at large times. 
The calculation of the one-loop meson self energy is done in a similar
way as for the fermion self-energy, i.e. by a Fourier-Bessel transform 
calculation of the convolution  momentum integral in
Eq. (\ref{se_ol}).
For zero meson momentum 
 the integration is done directly. 

We   have calculated the meson production in three ways. The first
possibility is
to take the Hartree initial state and the whole evolution of the
fermion Green's functions is in
the real times. For this case we calculate the meson production using 
Eq. (\ref{row_mes}) with only real time integration on the right hand side.
For the case of the correlated initial state the evolution 
of the fermion Green's functions is performed 
both on the real and imaginary part of the contour in
Fig. \ref{contour}. The corresponding meson production rate can be
calculated in two ways. One way is  to integrate only over real times
on the right hand side of Eq. (\ref{row_mes}), the other one is to
take the whole expression (\ref{row_mes}). Of course, at large times,
the two approaches give the same result, since the contribution of the 
imaginary time integration is negligible for large times. However, at
the beginning of the evolution, the differences are important. In most
of the cases, the full expression (\ref{row_mes}) is well behaving,
with less oscillatory
behavior. The oscillatory behavior at the beginning of the evolution is 
due to the finite time integration. This gives only on average an
 expression for the production rate. 

\begin{figure}
\begin{center}
\epsfig{file=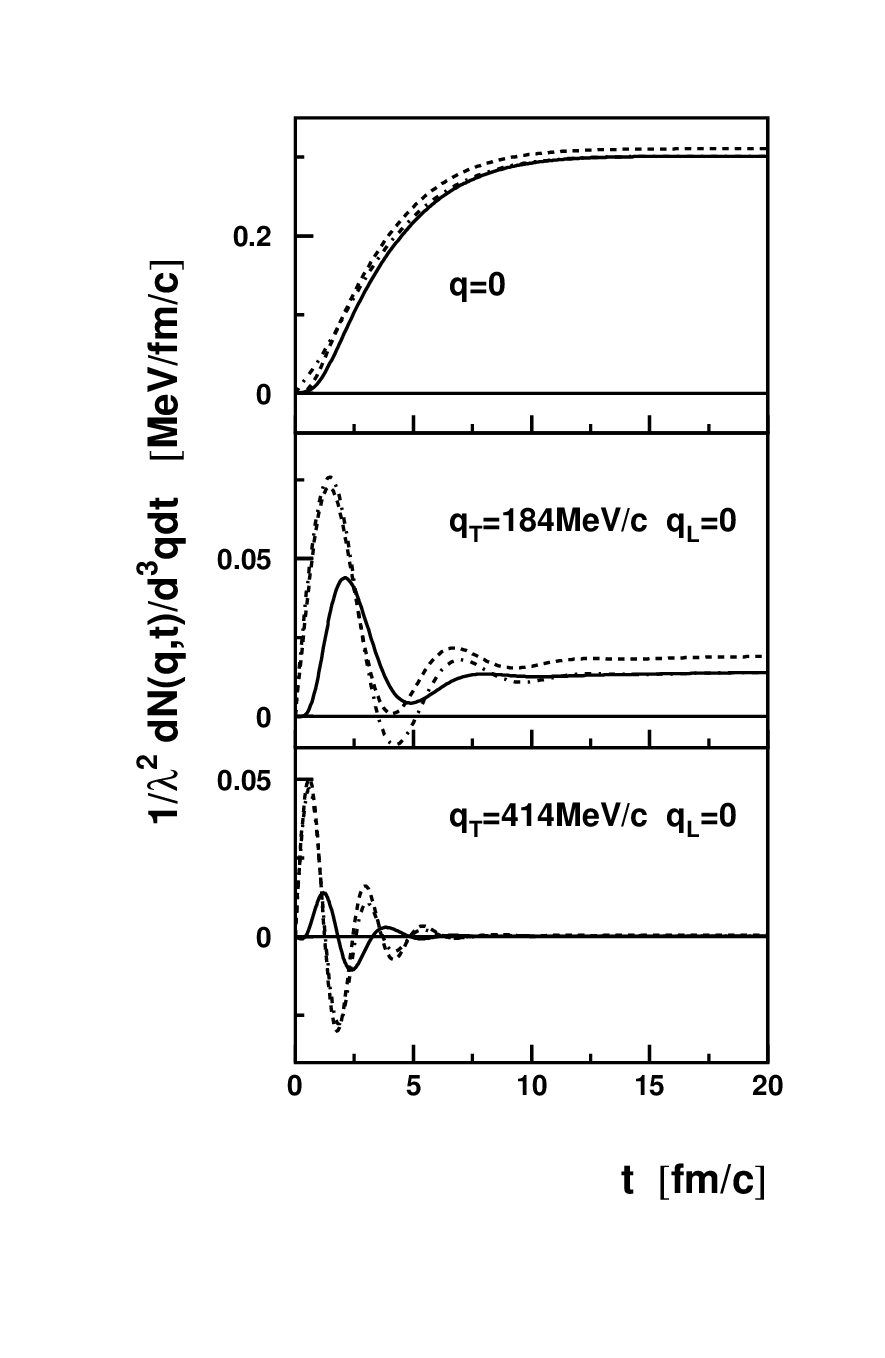,width=0.6\textwidth}
\vspace{0.5cm}
\end{center}
\caption{Quantum rate of the production of mesons  of mass $20MeV$ as a
function of time. The dashed line is the meson production rate for the 
 the uncorrelated initial state. The solid line is
the meson production rate for the correlated initial state with both
real and imaginary time integration in the meson production rate
 (\ref{row_mes}). The dashed-dotted line is the meson production
rate for the correlated initial state with only real time integration 
in the meson production rate
 (\ref{row_mes}). The upper, middle and lower panel 
correspond to the transverse momentum of the produced meson of $0$, $184$
and $414MeV/c$ respectively. The energy of the collision is $180MeV/$nucleon.}
\label{td20}
\end{figure}

\begin{figure}
\begin{center}
\epsfig{file=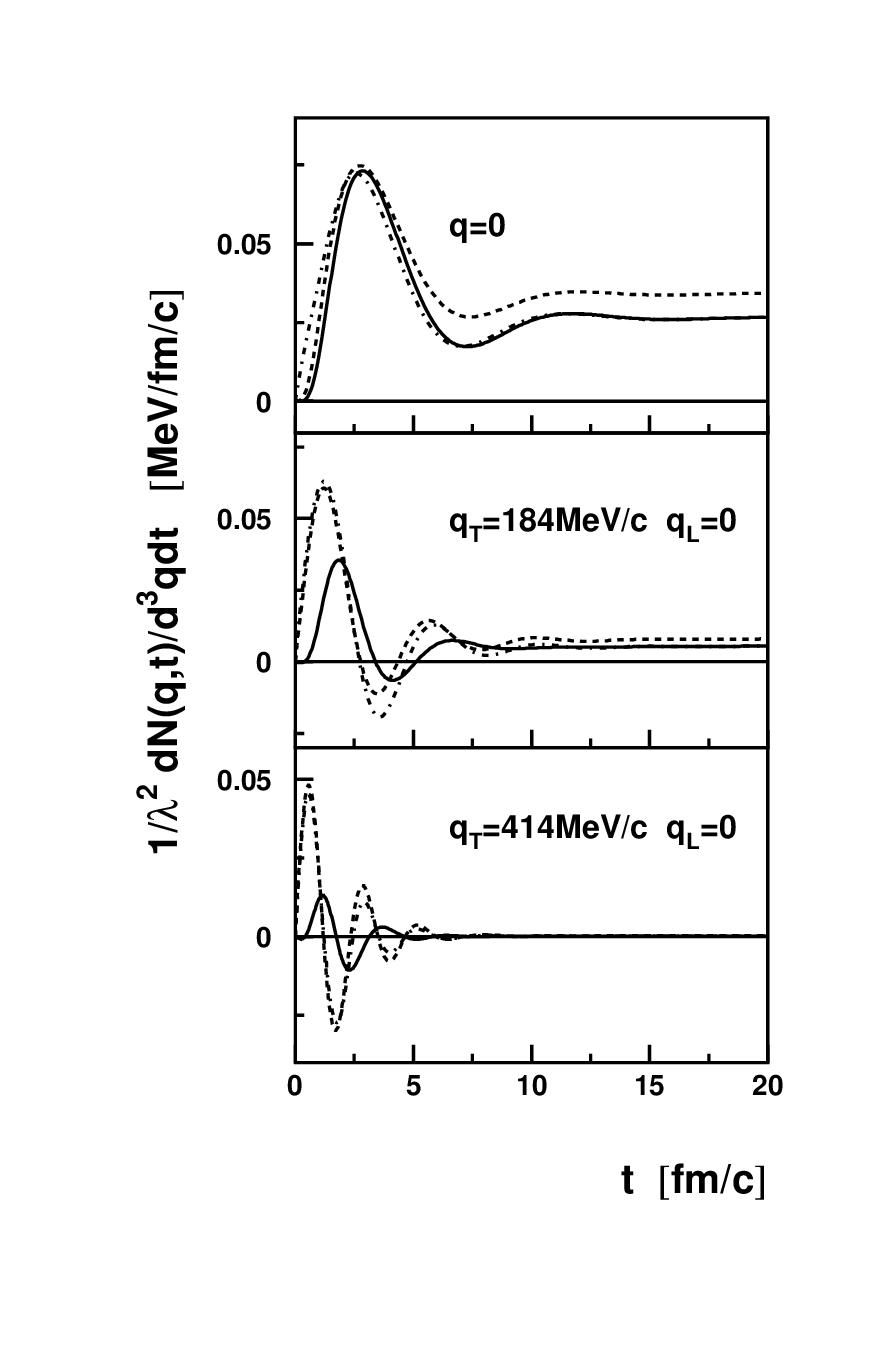,width=0.6\textwidth}
\vspace{0.5cm}
\end{center}
\caption{Quantum rate of the production of mesons of mass $140MeV$ as a
function of time. Details are similar as in Fig. \ref{td20}}
\label{td140}
\end{figure}

\begin{figure}
\begin{center}
\epsfig{file=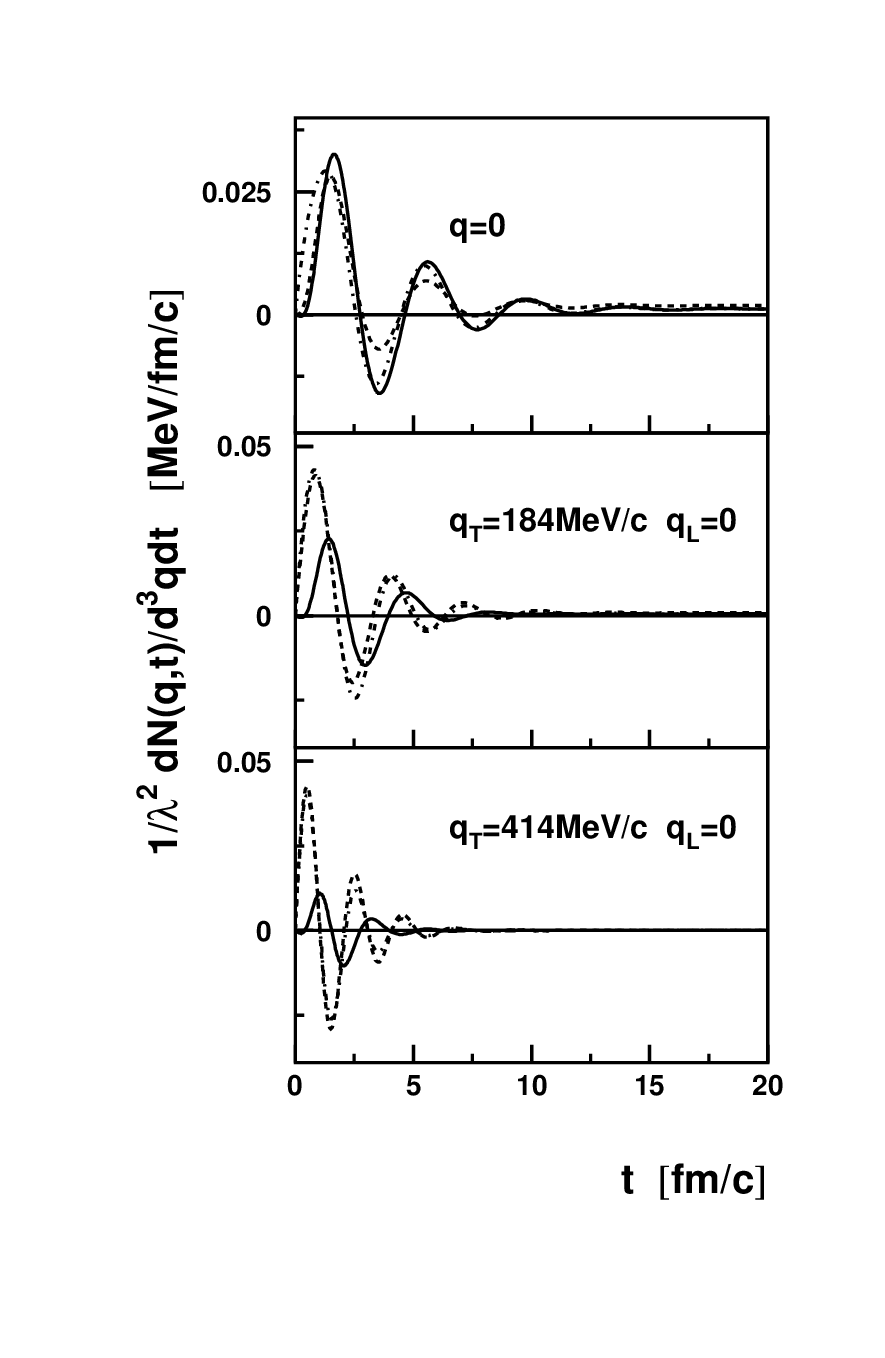,width=0.6\textwidth}
\vspace{0.5cm}
\end{center}
\caption{Quantum rate of the production of mesons of mass $300MeV$ as a
function of time. Details are similar as in Fig. \ref{td20}}
\label{td300}
\end{figure}

In Figs. \ref{td20}, \ref{td140} and \ref{td300} is  shown the time
dependence of the scaled production rate
$\frac{1}{\lambda^2}\frac{dN}{d^3qdt}$
 for three different masses of the
mesons $20$, $140$ and $300MeV$ respectively. 
We prefer to use the scaled production rate, since it allows us a
general discussion.
In particular, we can analyze the behavior around $q=0$ where the
production rate would be $0$ if multiplied by the pion vertex strength.
The general behavior is as expected, starting with oscillations until
the asymptotic production rate settles in. This initial time dependence 
of the production rate is not due to the nonequilibrium character of
the evolving system. Rather, it is due to the finite time integration in
the expression for the production rate, as discussed above.
For not very large values of the meson mass $m_s$ or meson momentum
$q$, i.e. not very far from the mass shell, we can parameterize the
fermion
Green's functions using a Lorentzian spectral function.
In that case one expects that the oscillations in the production rate
would be damped with a time $1/\Gamma$, where $\Gamma$ is the 
average width of the spectral function. In our case the width is around
$50MeV$
which should give a decay time of $4fm/c$ in accordance with the
Figs. \ref{td20}, \ref{td140} and \ref{td300}.
The asymptotic value of the production rate at large time is not
zero. However, for large mass of the meson or large momentum of the meson
the asymptotic value of the production rate is much smaller then the
amplitude of the oscillations at the initial time.
In order to get stable results for the case of large mass and/or
momentum of the produced meson
we had to use very small time step ($.067fm/c$ and $.15fm/c$ for the real 
and imaginary evolution respectively). This allows to get a correct
estimate of the fermion spectral function far from the mass-shell.
The asymptotic production rate of mesons is systematically larger for
the evolution from the uncorrelated initial state (dashed lines in
Figs. \ref{td20}, \ref{td140} and \ref{td300}). It is a manifestation
of the the higher temperature reached in the equilibration of the 
uncorrelated initial state. 
In Fig. \ref{tdlong} is shown the time dependence of the meson
production rate for large times. For the case of meson of  mass $140$
and $300MeV$ an 
almost stationary production rate is reached. However for the mesons
of mass $500MeV$ still oscillations of high amplitude are seen.
These oscillations are slowly damped to a positive value corresponding 
to the asymptotic meson production rate.

In Fig. \ref{tran_fig} is shown the production rate of mesons at large 
time ($t=20fm/c$), as a function of the transverse momentum. At this
time the production rate of mesons is almost isotropic.
The quantum meson production rate for large meson energy was
estimated as the average over its last oscillation in the time
dependent meson production rate.
 Similarly as
in the plots of the  time development of the production rate, the
production rate is larger for the uncorrelated initial state.
One should note that the plots in Fig. \ref{tran_fig} 
do not represent  the genuine transverse momentum
distribution of produced mesons, since the interaction strength 
$\lambda^2$ can in general depend quite strongly on the momentum of the
produced meson $q$.
In Fig. \ref{trancal_fig} is shown the total number of mesons
produced during the evolution as 
a function of the transverse momentum. 
The highest number of mesons is produced in the evolution from
an uncorrelated initial state. In the time integrated production rate, 
a big difference appears between the meson production calculated from
the full expression 
 (\ref{row_mes}) and  the  meson production calculated from
the Eq. (\ref{row_mes}) including only the real interval of the time
integration. The difference can be observed in the plots of the time
dependence of the meson production rate for small times.
The large oscillations of the production rate at small times give after 
time integration a large contribution.
This contribution is unphysical, since in a nuclear collision there is no
sudden switching on of the interaction at time $t=t_0$. The time
integrations are always infinite and the production rate should not
have the oscillations characteristic of the finite time integration. 
At each time the locally asymptotic, but not necessarily stationary,
meson production rate 
can be defined, if the time integration would be much larger than the 
 inverse fermion single particle width $1/\Gamma$.
In the nonequilibrium Green's
evolution presented in this work, we have to start  from a finite
initial time. The procedure with imaginary time integration in the
initial state leads to a better description of the evolution of fermions
and  of the  meson production rate. 

Close to the mass-shell the fermion Green's functions are often
parameterized as~:
\begin{eqnarray}
\label{lopa}
G^{<}(p,t_1,t_2) &=&i f(p) \exp \left(-i\omega_p(t_1-t_2)
 -\Gamma_p|t_1-t_2| \right) \ ,
\nonumber \\
G^{<}(p,t_1,t_2) &=&-i (1-f(p)) \exp \left(-i\omega_p(t_1-t_2) 
 -\Gamma_p|t_1-t_2|  \right) 
\ ,
\end{eqnarray}
and one gets for the meson production rate~:
\begin{equation}
\label{innol}
\frac{dN(q=0,t)}{d^3qdt} = \lambda^2 \int \frac{d^3p}{(2\pi)^3}   
f(p)(1-f(p)) \frac{4\Gamma_p}{m_s^2+\Gamma_p^2} \ .
\end{equation}
This formula gives Lorentzian dependence on the meson mass if 
the width $\Gamma_p$ is weakly depending on momentum.
In Fig \ref{lor} is shown the dependence of the asymptotic production
rate of mesons at zero momentum as a function of the meson mass.
This dependence is close to an exponential and cannot be approximated
by a Lorentzian.
In the appendix \ref{eqpr} is shown that even if the fermion spectral function 
is Lorentzian we do not get a Lorentzian dependence of the meson
production rate on the meson mass, if we use a more general expression 
for the equilibrium Green's functions than the Eq. (\ref{lopa}).

The  fermion Green's functions obtained as a result of the numerical
solution of the Kadanoff-Baym equations 
are very
different from the parameterization (\ref{lopa}) \cite{d2}.
The function  $f(p)$ in the above parameterization
corresponds to the momentum distribution of fermions.
However, we have found that the imaginary part 
of $-iG^{<(>)}(p,t_1,t_2) \exp ( i \omega_p (t_1-t_2))$ is nonzero and
gives an important contribution to the meson production rate.
The Lorentzian parameterization of the spectral function is expected 
to be valid near the mass shell. Accordingly, it cannot be
applied for the description of the meson production which involves 
the far from the mass-shell spectral function of fermions.

\section{Comparison to the semiclassical production rate}

 The  semiclassical meson production rate
from a homogeneous BUU-like evolution of the collision will be presented.
The BUU evolution is done for the Hartree initial momentum
distribution and for the  initial momentum distribution obtained
from the imaginary time evolution of the Kadanoff-Baym equations.
Such calculations correspond to the standard approach to the meson
production in nuclear collisions.

Other results will also be presented for the semiclassical production rate
with fermion  momentum distribution taken from the equal time Green's
functions generated in the quantum evolution.
It is interesting to compare
the semiclassical meson production rate in a quasistationary state 
to the quantum production rate in the same
system.

The semiclassical production rate is defined in App. \ref{app_sem}.
In this chapter we use the expression (\ref{sc_og}) and (\ref{mael}),
with $\gamma=50MeV$.

\subsection{Semiclassical evolution equations and meson production}
\label{sem1}

The semiclassical evolution equation for fermion densities can be obtained 
from the Kadanoff-Baym equations \cite{kb,d1,ma}. We take the
nucleon-nucleon
scattering cross section corresponding to the direct Born fermion self 
energy~:
\begin{eqnarray}
\label{sc_bo}
 \frac{df(p_1,t)}{dt} &  = &
-  \int \frac{d^3p_2}{(2\pi)^3}    
 \int \frac{d^3p_3}{(2\pi)^3}\int \frac{d^3p_4}{(2\pi)^3}
 V^2(p_1-p_3)  \nonumber \\ & & 
 (2\pi)^3 \delta^3(p_1+p_2-p_3-p_4) 2\pi \delta(\omega_{p_1}+
\omega_{p_2}-\omega_{p_3}-\omega_{p_4}) \nonumber \\ & & 
\Bigg(
f(p_1,t) f(p_2,t)\left( 1 -f(p_3,t)\right)\left( 1 -f(p_4,t)\right)
\nonumber \\ & & - 
f(p_3,t) f(p_4,t)\left( 1 -f(p_1,t)\right)\left( 1 -f(p_2,t)\right)
\Bigg) \ .
\end{eqnarray}
The initial momentum distribution of fermions $f(p,t_0)$, 
should be taken as two Fermi spheres separated in momentum space by the
 momentum of the collision. However, we know that the nuclei in the
ground state are not Hartree states with with zero temperature.
In particular they have a large momentum component in the fermion
momentum distribution. In our model involving spatially homogeneous
system, it can be described by taking as the initial state the 
Hartree state evolved on the imaginary time contour by the
Kadanoff-Baym equations.
This corresponds to allowing for the initial quantum
correlations, which modify the momentum distribution. This effect is
expected to be important for the subthreshold meson production rate.
Accordingly, we will use also the fermion momentum distribution in the 
correlated state as the initial momentum distribution $f(p,t_0)$.
 The average kinetic energy
in the correlated state is around $10MeV/$nucleon larger than in
the Hartree state. 
The momentum distribution in the correlated initial 
state has  a large momentum component.

The BUU-like evolution  following the semiclassical fermion collision
term (\ref{sc_bo}) was solved on the same momentum grid as the
quantum evolution. The time step was taken equal to $0.5fm/c$.
In the effectively 5 dimensional integration on the right hand
of Eq. (\ref{sc_bo}) 3 dimensions were integrated by Monte-Carlo integration
and two momentum integration were summed over the momentum grid.

 From the semiclassical fermion
momentum distribution obtained from Eq. (\ref{sc_bo}), we can calculate 
the semiclassical meson production rate using Eq. (\ref{sc_og}).
Another possibility is to use the fermion momentum distributions
obtained from the quantum evolution equations~:
\begin{equation}
f(p,t)=-i G^<(p,t,t) \ .
\end{equation}
As the system in the quantum evolution reaches an almost stationary
state after $20fm/c$ of real time evolution, we can  
compare the meson production rate in this stationary state as obtained
from the quantum and the semiclassical expressions for the meson
production.
The two stationary states that we discuss below have different
temperatures and correspond to the initial correlated state and the initial
Hartree state in the quantum evolution equations.

\subsection{Discussion of the results for the semiclassical meson
production rate}

The time dependence of the semiclassical production rate is very
similar for different
meson masses or momenta. Thus, we present only one Figure of
the time dependence of the semiclassical production rate. In Fig. 
\ref{cltd} is shown the time dependence of the semiclassical meson
production rate for the meson momentum $q=184MeV/c$ and for three
different meson masses.  At the initial time the semiclassical
production rate has two values, depending on the initial state of the
evolution. For the correlated initial state (dotted and dashed-dotted
lines in Fig. \ref{cltd}) the production rate at the beginning
of the evolution  is higher than   for the Hartree initial state
(solid and dashed lines in Fig. \ref{cltd}). It is due to the fact
that the average kinetic energy in  the correlated initial state
is higher than in the Hartree state.

 The BUU-like evolution from the 
initial Hartree state, has the lowest average kinetic energy. 
For the higher energies of the produced mesons, especially for the case
of $ \Omega_q=352MeV$ (the lower panel), the semiclassical production
rate increases  during the evolution. Such a behavior is
expected, since the system approaches thermal equilibrium with some 
small thermal high momentum tail. However, as the average kinetic energy is
conserved, the final equilibrium state has the lowest
semiclassical meson production rate among all the results presented in
Fig. \ref{cltd}.

On the other hand, the quantum transport evolution starting from the
initial Hartree state, leads to an increase of the kinetic energy.
The initial correlation energy is changed into the kinetic energy
\cite{d2,koh}. As a result the average kinetic energy is increased by 
$30 MeV$/nucleon and it is the largest among all the different
evolutions considered in Fig. \ref{cltd}.
Thus, it is obvious that also the semiclassical production rate would
be the largest in this case. Within the first few $fm/c$ the
semiclassical production rate increases very fast. 
After this initial increase the meson production rate increases only
slowly.
This slow change of the semiclassical meson production rate is
determined by the equilibration of the momentum distribution of
fermions in the later stage of the Kadanoff-Baym equations evolution 
of fermions.

The two evolutions starting from the correlated initial state 
have very similar
average kinetic energies during the whole evolution. Thus, the
semiclassical meson production rate for the BUU-like evolution and 
for the quantum evolution from a correlated initial state are not very
different (the dotted and dashed-dotted lines in Fig. \ref{cltd}).
 The BUU-like evolution starting from the correlated initial 
state can be considered as a model of the BUU evolution of the initial
nucleon momentum distributions in nuclei that includes the large
momentum tail
of quantum origin (dotted line in Fig. \ref{cltd}). This kind of calculation
 represents all the quantum effects that can be 
taken into account in the approach 
using the 2-nucleon process only and on-shell particles.
To go beyond  the discussion of the quantum effects in the initial state only,
i.e. to include the off-shellness of
the fermion Green's functions, requires an off-shell calculation of the
meson production rate as was presented in Sect. \ref{mp} and
\ref{nmp}. 
The difference between the BUU evolution from the Hartree initial
state and the BUU evolution from the correlated initial state is
especially important for the case of the production of meson of mass 
$300MeV$. This meson mass corresponds to the subthreshold meson
production, since the energy of the produced meson is $352MeV$ and the 
average c.m. energy is $180MeV$ (the kinematical limit for the meson 
production in the Hartree initial state is around $450MeV$).
The ratio of the semiclassical meson production in the correlated
initial state evolution and in the Hartree initial state evolution 
changes from $26$ at $t=0$ to $3$ at $t=20fm/c$~.
From Fig. \ref{cltd} we see that the BUU-like evolution gives
similar results as the the quantum evolution if the correlated initial 
state is used. The use of the initial correlated state is justified in 
the Kadanoff-Baym equations. It is a much more realistic way of starting 
the quantum transport equations evolution than the Hartree initial state.

We do not show in Fig. \ref{cltd} the quantum meson production rate
since the quantum rate is  one order of magnitude larger.
Moreover, the asymptotic quantum meson production rate can be defined
only for large times. The large time quantum and semiclassical
meson production rate as a function of the transverse momentum of the
produced meson are compared in Fig. \ref{cltr}.
The quantum meson production rate is about one order of magnitude
larger than the semiclassical meson production rate.
The dashed and the dotted line represent the quantum and semiclassical 
meson production rate for the larger temperature state reached in the
evolution from the Hartree initial state. The solid and the dashed-dotted
line represent the same for the evolution from the correlated initial state.
From Fig. \ref{cltr} we see that the semiclassical and the quantum meson
production rates are very different for the same state reached in the
evolution of the system.
This is very different from Fig. \ref{cltd}, where we have seen that
the differences between the semiclassical and quantum evolutions 
of the correlated initial state are not big if we are using
only the semiclassical meson production rate formula.

In Fig. \ref{tr14} we present the transverse momentum distribution of
the meson production rate at $t=20fm/c$
 for the lower energy of the collision ($110MeV/$nucleon in 
the c.m.). The three ``physically'' well founded approaches are
the quantum evolution from the correlated initial state, the BUU
evolution from the correlated initial state \cite{com} 
and the BUU evolution from the Hartree initial 
 state  (the solid, dashed-dotted and dotted lines
in Fig. \ref{tr14}). Besides these results we plot the semiclassical
meson production rate for the fermion momentum distribution calculated
in the quantum evolution of the correlated initial state. Similarly
as for the higher energy the semiclassical meson production rate is similar 
for the BUU and quantum evolved correlated initial state.
Also similarly as in the results presented in Fig. \ref{cltr}
 the quantum meson production rate is about one
order of magnitude larger than the semiclassical meson production
rate. The difference is even larger if compared to the semiclassical
meson production rate for the BUU evolution of the Hartree state.
We have used the same effective width $\gamma=50MeV$ in the
calculation of the matrix element for  
semiclassical meson production, as for the higher 
energy, since the single particle fermion width is reduced only by a
few percent, when changing the initial energy.
 We do not show the results for $m_s=300MeV$, because no 
asymptotic meson production rate could be estimated for this meson
mass and large meson momentum. 

\section{Conclusions}

We have presented the first approach to the
calculation of the meson production rate in the non-equilibrium
quantum transport equations.
The Kadanoff-Baym equations for the evolution of fermions were solved
and the resulting off-shell fermion Green's functions were used for
the calculation of the one-loop meson self-energy.
The one-loop meson production rate shows strong oscillations at the
beginning
of the evolution of the system. This behavior comes from the fact that the 
time integral in the meson evolution equation is finite.
Only after some time the meson production rate reaches some asymptotic 
value, which can be defined as the quantum meson production rate.
We have found that the behavior of the meson production rate at small
times improves if the imaginary part of the time integration in the
meson evolution equations is taken into account.
We plan to study this question in  details for the case
where the fermion Green's functions 
 on the imaginary and real parts of the time contour 
is the finite temperature Green's function corresponding to the same
stationary finite temperature state.

The formalism based on the off-shell nucleon propagators allows to
estimate the effects of the initial quantum correlations on the 
meson production rate. We have found that these initial correlations
are
important and that their inclusion  increases the semiclassical meson 
production rate. This effect is especially important for the
subthreshold meson production ($m_s=300MeV$ in Fig. \ref{cltd}).
The initial correlations in the fermion
distribution
 can be evolved with
BUU-like fermion evolution equation.
Also the fermion momentum distribution  can be obtained from the
equal time fermion Green's functions calculated with the Kadanoff-Baym 
equations.
Both calculations give similar average fermion kinetic energy and
similar semiclassical
meson production rate. Thus, it makes no dramatic difference in the
semiclassical particle production if the nucleon momentum
distribution is evolved according to the Kadanoff-Baym equations or
according to the BUU equations, if the initial momentum distribution
includes the quantum correlations. 

The situation is very different if the quantum meson production rate
is used to calculate the number of produced particles.
In the system
reached by the equilibration of the nuclear collision of the energy
$180MeV/$nucleon in c.m,
the quantum meson production rate is usually one order of magnitude
larger then the semiclassical meson production rate.
This is an important difference which requires detailed attention when
discussing the particle production in a hot nuclear matter.
The phenomenological consequences for the pion production may be less 
important since we discuss only the direct pion production
process. The direct pion production is known to be less important than 
the 
delta production, at least in the semiclassical approach. The delta 
production should be less influenced by the nucleon off-shellness.
However, as a result of our estimates
 the direct pion production should not be
neglected in the total pion production yield if the quantum 
production rate is used for the direct pion production.
More detailed calculations using a model incorporating delta,
nucleon and pion degrees of freedom is required to address
quantitatively the phenomenological predictions for  pion production.
More work is also needed in the study of the parameterization
of the retarded nucleon propagator which comes into the calculation of 
the semiclassical meson production rate. This could be studied first in the
equilibrium state using the quantum transport equations.

In view of our results,
the models of the particle production in the hot nuclear medium
should be based on the Kadanoff-Baym equations, or should be based on
off-shell
transport equations \cite{he_pr}. 
Great care must be shown in the use of approximate fermion spectral functions.
The numerical calculations show that the simple parameterizations
of the fermion spectral function as  in Eq. \ref{onloe}
does not give the correct quantitative result.
 This is even more pronounced  for the less general
parameterization of the fermion Green's function in Eq. (\ref{lopa}),
giving a qualitatively different dependence on the  meson mass.

 \acknowledgments
The author would like to thank
P. Danielewicz for useful discussion.
 Also, he gratefully acknowledges the hospitality extended to
him by the YITP.

\appendix

\section{Semiclassical production rates}
\label{app_sem}

The semiclassical production rate can be obtained from the 
cut diagrams, by putting the Greens functions $G^{<(>)}$ on shell
\cite{d1,ma,d3}.
This procedure leads  to zero for the one-loop diagram
 (Fig. \ref{one_loo}) that was discussed for the quantum production
rate.
The 3-body energy conserving delta function has no solution for our
kinetic conditions.
In the next order in $V$, only a vertex correction to the one-loop
diagram appears. This also does not give any contribution on-shell.
In the second order in the interaction $V$ there appear $8$ direct diagrams
(Figs. \ref{diag_se}, \ref{diag_inne} and \ref{diag_ze}). There are 8
other exchange diagrams, that  we will not discuss here since 
for the fermion evolution only direct diagrams are taken.
 The diagrams in Fig. \ref{diag_se} should not appear in the
perturbative expansion, since  they represent self-energy corrections
on fermion lines. These are already included in the dynamically
calculated
fermion Green's functions. However, they should appear naturally in 
a perturbative expansion of the production rate in vacuum, using
the on-shell Green's functions. 
The separation of the 3-body event which is the 
scattering with meson production from 
the succession of a 2-body scattering and successive meson emission
can be done by the non-zero width in the retarded propagator in the
diagrams
in Figs. \ref{diag_se}, \ref{diag_inne} and \ref{diag_ze}
 \cite{d3}.
 Of course the 2-body scattering and successive meson
emission gives zero contribution on shell. 
 If such a procedure is used, the diagrams in
Fig. \ref{diag_se} are a first term in a series of diagrams involving
the direct Born self-energy insertions on fermion lines.
The sum of the diagrams in such a series should correspond to the
one-loop meson-self energy diagram with resumed fermion Green's functions.
One diagram of the next order in this series is shown in Fig. 16.
%\ref{seriad}.
The cutting procedure on the higher diagrams in this  series leads to
the processes involving many fermions on-shell. Such many nucleon
processes should be  taken into account in the calculation of the
subthreshold particle production. The many nucleon processes, which can be
obtained from the cutting of the direct Born self
energy insertions,  are   resumed in the off-shell Green's functions
calculated
in Sect. \ref{qte}. In the following we will take only the diagrams in 
Fig. \ref{diag_se} for the calculation of the semiclassical meson
production rate. This correspond to taking only the 2-nucleon term
in the expansion of the meson self energy
 in the number on interacting nucleons.
The corresponding matrix elements is given below (\ref{mael}) (Fig. 
\ref{mat_se}).
The diagrams in Fig. \ref{diag_ze} are zero in the case of the interaction
vertex (\ref{inteham}) and also more generally for a large class of
theories. 

  Identifying each cut 
propagator with the fermion Green's function 
on shell ($i G^{<(>)}$ depending on the direction of
crossing the cut), and the propagator on the left (right) side of the
cut with the retarded (advanced) propagator we obtain the following
form of the meson production rate~:
\begin{eqnarray}
\label{sc_og}
\frac{dN(q,t)}{d^3qdt} &  = & 
\int \frac{d^3p_1}{(2\pi)^3} \int \frac{d^3p_2}{(2\pi)^3}    
 \int \frac{d^3p_3}{(2\pi)^3}\int \frac{d^3p_4}{(2\pi)^3}
|M|^2 (2\pi)^3 \delta^3(p_1+p_2-p_3-p_4-q) \nonumber \\
& & 2\pi \delta(\omega_{p_1}+
\omega_{p_2}-\omega_{p_3}-\omega_{p_4}-\Omega_q) 
f(p_1,t) f(p_2,t)\left( 1 -f(p_3,t)\right)\left( 1 -f(p_4,t)\right) \ .
\end{eqnarray}
Here analogously as for the calculation of the quantum production rate 
(sect. \ref{mp}) we put the meson Green's function on
shell and neglect the meson occupancy in the 
calculation of the production rate.
The square of the matrix element $|M|^2$ in the above expression can
be identified with the 
sum of the cut meson self-energy diagrams (Fig. \ref{diag_se},
\ref{diag_inne} and \ref{diag_ze}) with the 
cut fermion Green's functions $i G^{<(>)}$ factored out.
The square of the matrix element $|M|^2$ can be written in the form of 
the diagram in Fig. \ref{mat_fu} where the fermion Green's function is the
retarded propagator and  the external Green's functions have 
momenta  corresponding to the integration momenta in
Eq. (\ref{sc_og}).
The diagrams in Fig. \ref{diag_inne} and \ref{diag_ze}
 appear also in the expansion of the self-energy in the
quantum transport equations,
i.e. in the expansion using the resumed fermion propagators, but at a
higher order in $V$.
Thus in order to compare the semiclassical rate to the one-loop
quantum rate,  only the cut self-energy insertion
diagrams (Fig.  \ref{diag_se}) should be taken.
Clearly, even using the finite width for the retarded propagator in
the
semiclassical calculation no correspondence is found between the order 
of the perturbative expansion on shell and off shell. The cut
diagrams in Fig.  \ref{diag_inne} and \ref{diag_ze}
 give contributions of the same order as the 
  cut self-energy insertion diagram (Fig.  \ref{diag_se}).
However, in the quantum expansion theses diagrams  are of different order.
Moreover, already at the first order in $V$ a vertex correction to the 
one-loop quantum diagram appears, which is still 
zero on-shell.

For the diagrams in Fig. \ref{diag_se}
  we obtain (Fig \ref{mat_se})~:
\begin{eqnarray}
\label{mael}
|M|^2  = & 4 \lambda^2 V(p_2-p_4)^2 \Bigg( &
|\frac{1}{\omega_{p_1-q}+\omega_{p_2}-\omega_{p_3}
-\omega_{p_4}+ i \gamma}|^2 \nonumber \\
& &+|\frac{1}{\omega_{p_1}+\omega_{p_2}+\omega_{p_3+q}
-\omega_{p_4}+i\gamma}|^2 \Bigg) \ .
\end{eqnarray}
After the calculation of the traces over spin and isospin indices, the 
interaction $g^2$ and a factor $4q^2$ appear. This can be 
written in the same way as for the one-loop calculation, i.e. as $\lambda^2$.
The additional factor $4$ comes  from the spin isospin trace in the
second fermion loop.
In most of the cases, unless otherwise stated,
 we use the production rate with constant  width
parameter $\gamma=50MeV$. This is the typical average fermion width 
as obtained from  
the quantum transport equations for fermions at the collision energies
considered here.
 Of course in general it is momentum
and time dependent, but in the calculation of the semiclassical
production cross section we take a constant value of the fermion width.

\section{Meson production in equilibrium}
\label{eqpr}

In this section we will present a simple estimate of the meson
production in an equilibrium fermion system at finite temperature.
For the fermion spectral function we will take the Lorentzian form.
The fermion Green's functions, after 
Fourier transforming
in the relative time variable, will be parameterized in the following
way~\cite{kb}~:
\begin{eqnarray}
\label{onloe}
G^<(p,\omega)=i
\frac{\Gamma_p}{(\omega-\omega_p)^2+(\Gamma_p/2)^2}f_0(\omega) \ ,
\nonumber \\
G^>(p,\omega)=-i
\frac{\Gamma_p}{(\omega-\omega_p)^2+(\Gamma_p/2)^2}(1-f_0(\omega)) \ ,
\end{eqnarray}
where we take a constant width $\Gamma_p$ and
\begin{equation}
f_0(\omega)=\frac{1}{1+exp\big((\omega-\mu)/T\big)} \ 
\end{equation}
is the equilibrium Fermi distribution at finite temperature
($T=30MeV$, $\mu=40MeV$, $\Gamma=60MeV$, $|p|< 900MeV$,
$|\omega-\omega_p|<1800MeV$).

We have calculated the meson production from the one loop diagram~:
\begin{equation}
\label{ol_eq}
\frac{dN(q=0)}{d^3qdt}=  \lambda^2\int \frac{d^3p}{(2\pi)^3}
\int \frac{d\omega}{(2\pi)} G^<(p,\omega)G^>(p,\omega-M) \ 
\end{equation}
The above equation is  different from the Eq. (\ref{innol})
used in Sect. \ref{nmp}. Here we are using the general
equilibrium form of the Green's functions \cite{kb} with a Lorentzian
spectral function. As a result, the dependence of the meson production
rate on the meson mass is almost exponential (see
Fig. \ref{proste_fig}),
 similarly as in the
numerical results in Fig. \ref{lor}. This dependence is not
Lorentzian as expected from Eq. (\ref{innol}).
 Even for Lorentzian spectral function with constant
width the 
resulting dependence of the production rate on the meson mass is 
not Lorentzian. This makes impossible a direct estimation of the
single particle fermion width from the comparison of the meson production 
rate for different meson masses.

In Fig. \ref{proste_fig} is also shown the semiclassical meson
production rate calculated from the Eq. (\ref{sc_og}) at zero meson
momentum.
For the equilibrium fermion momentum distribution we have taken
the equilibrium momentum distribution at finite temperature~:
\begin{equation}
f(p)=f_0\left(\frac{p^2}{2m}\right)
\end{equation}
or the momentum distribution including the quantum effect
\begin{equation}
\label{lata}
f(p)=-i\int\frac{d\omega}{2\pi}G^<(p,\omega) \ .
\end{equation}
In the second case the rate of the meson production is much larger,
reflecting
the large momentum tail in the momentum distribution calculated 
 from Eq. (\ref{lata}).
Similar large momentum tail is present in nuclear matter at zero
temperature. This effect was found to be important for the semiclassical
meson production in our nuclear collision model (Sect. \ref{nmp}).
In Fig. \ref{proste_fig} are shown also the semiclassical meson
production rates for the width parameter $\gamma\rightarrow 0$.
This strongly modifies the meson production rate for small meson mass.
The meson production rate has a singularity for $m_s \rightarrow 0$.
However, the nonzero $\gamma$ parameter changes this behavior. The
meson production rate stays finite for $m_s \rightarrow 0$ if 
$\gamma>0$. This is similar to the Landau-Pomeranchuk-Migdal effect
discussed in \cite{kn_vo}. 
 The semiclassical production cross section in medium
is modified and has no singularity, even-though the meson production is
given by the  semiclassical expression  (\ref{sc_og}).

The quantum meson production rate is  much larger than the
semiclassical production rate in Fig. \ref{lor}. 
However, such a comparison is not very meaningful.
The Green's functions $G^{<(>)}$ used in the one-loop
meson self energy (\ref{onloe}), where not calculated
selfconsistently for a system of fermions interacting with potential
$V$. On the other hand the interaction potential $V$ enters in the
expression for the semiclassical production rate of mesons.

\vfill

\newpage

\newpage

\newpage

\newpage

\newpage

\newpage
\begin{figure}
\begin{center}
\epsfig{file=tdlong.eps,width=0.6\textwidth}
\vspace{0.5cm}
\end{center}
\caption{Quantum rate of the production of mesons at zero momentum as a
function of time. The dashed line is the meson production rate for the 
 the uncorrelated initial state. The solid line is
the meson production rate for the correlated initial state with both
real and imaginary time integration in the meson production rate
 (\ref{row_mes}). The dashed-dotted line is the meson production
rate for the correlated initial state with only real time integration 
in the meson production rate
 (\ref{row_mes}). The upper, middle and lower panel 
correspond to the mass of the produced meson of $140$, $300$
and $500MeV$ respectively. The energy of the collision is $180MeV/$nucleon.}
\label{tdlong}
\end{figure}

\newpage
\begin{figure}
\begin{center}
\epsfig{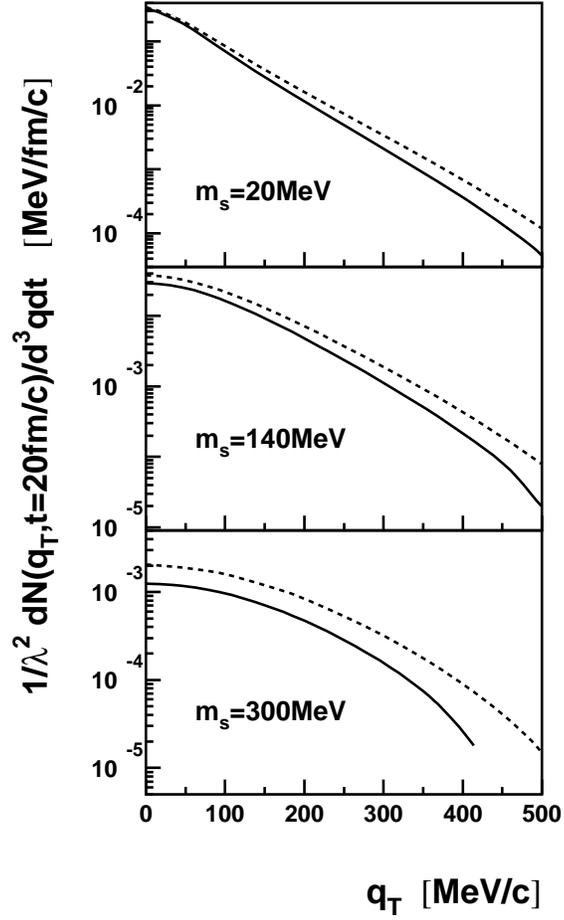}
\vspace{0.5cm}
\end{center}
\caption{Quantum rate of the production of mesons as a
function of the transverse momentum at the end of the evolution ($t=20fm/c$).
 The dashed line is the meson production rate for 
 the uncorrelated initial state. The solid line is
the meson production rate for the correlated initial state. 
  The upper, middle and lower panel 
correspond to the mass of the produced meson of $20$, $140$
and $300MeV$ respectively. The energy of the collision is $180MeV/$nucleon.}
\label{tran_fig}
\end{figure}

\newpage
\begin{figure}
\begin{center}
\epsfig{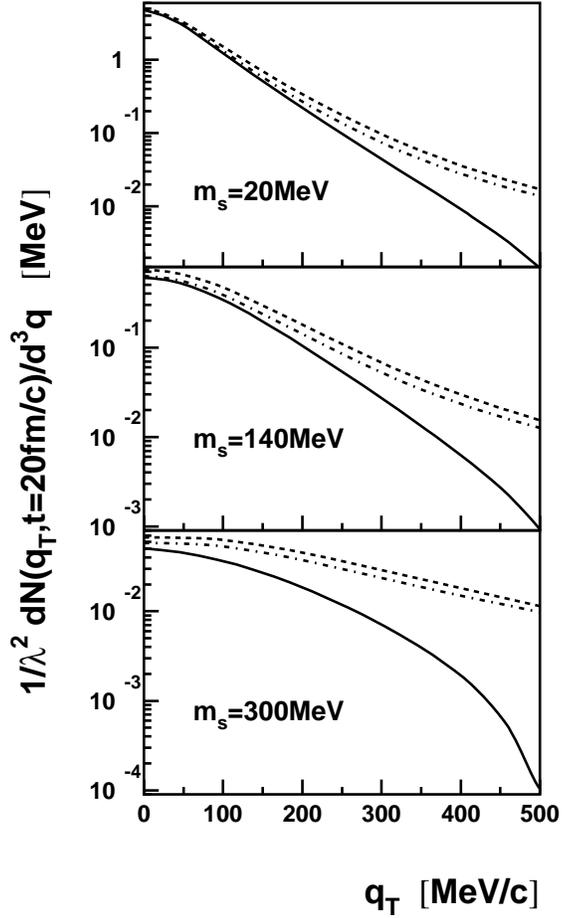}
\vspace{0.5cm}
\end{center}
\caption{The number of produced  mesons as a
function of the transverse momentum at the end of the evolution
($t=20fm/c$)
 for the quantum one-loop meson production rate.
 The dashed line is the number of produced mesons  for the 
 the uncorrelated initial state. The solid line is
the number of produced  mesons  for the correlated initial state
with both real and imaginary time integration in the meson production rate
 (\ref{row_mes}). The dashed-dotted line is the number of produced 
mesons for the correlated initial state with only the real time
integration in the meson production rate (\ref{row_mes}).
  The upper, middle and lower panel 
correspond to the mass of the produced meson of $20$, $140$
and $300MeV$ respectively. The energy of the collision is $180MeV/$nucleon.}
\label{trancal_fig}
\end{figure}

\newpage
\begin{figure}
\begin{center}
\epsfig{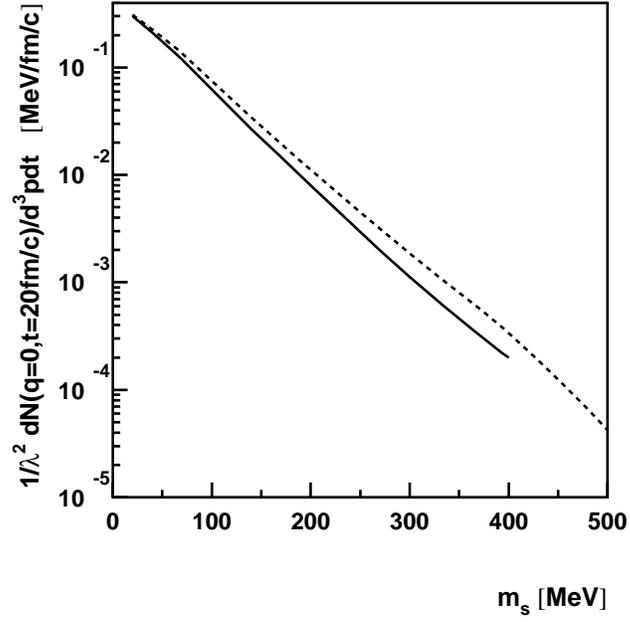}
\vspace{0.5cm}
\end{center}
\caption{Quantum  rate of the production of mesons as a function of the
meson mass at the end of the evolution ($t=20fm/c$), for mesons with
zero momentum. The dashed line is the meson production rate for the
uncorrelated initial state.
The solid line is the meson production rate for the correlated initial 
state. The energy of the collision is $180MeV/$nucleon.}
\label{lor}
\end{figure}

%\newpage
%\begin{figure}
%\begin{center}
%\epsfig{file=hist.eps,width=0.7\textwidth}
%\vspace{0.5cm}
%\end{center}
%\caption{The contour plots of the 
%initial fermion  momentum distribution for the correlated initial
%state (upper panel) and for the usual Hartree initial state
%(lower panel).}
%\label{hist_fig}
%\end{figure}

\newpage
\begin{figure}
\begin{center}
\epsfig{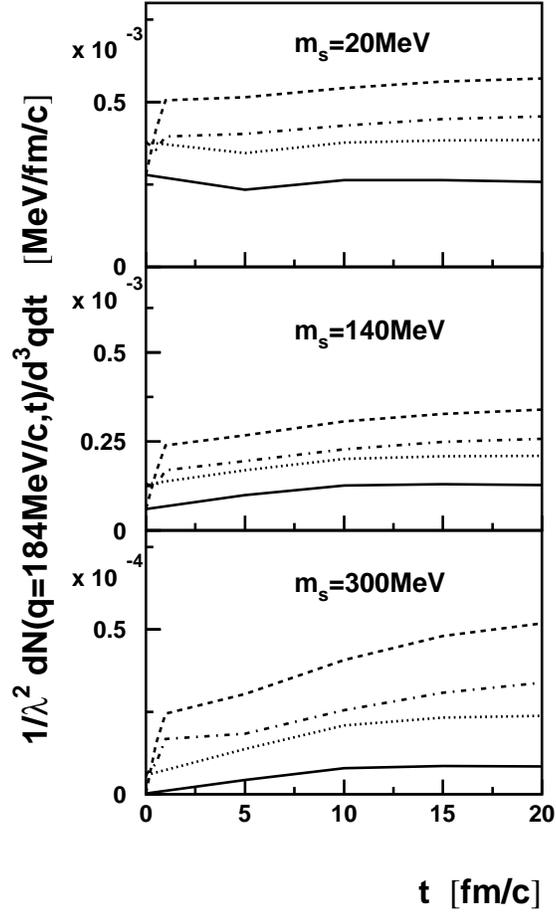}
\vspace{0.5cm}
\end{center}
\caption{Semiclassical
rate of the production of mesons at  momentum $q=184MeV/c$  as a
function of time. 
The solid and the dotted lines represent the semiclassical meson
production rate for the BUU-like evolution of the fermion distribution
from a Hartree and a correlated initial state respectively.
The dashed and the dashed-dotted lines represent the semiclassical
 meson production rate for the quantum transport equations
 evolution of the fermion distribution
from a Hartree and a correlated initial state respectively.
 The upper, middle and lower panel 
correspond to the mass of the produced meson of $140$, $300$
and $300MeV$ respectively. The energy of the collision is $180MeV/$nucleon.}
\label{cltd}
\end{figure}

\newpage
\begin{figure}
\begin{center}
\epsfig{file=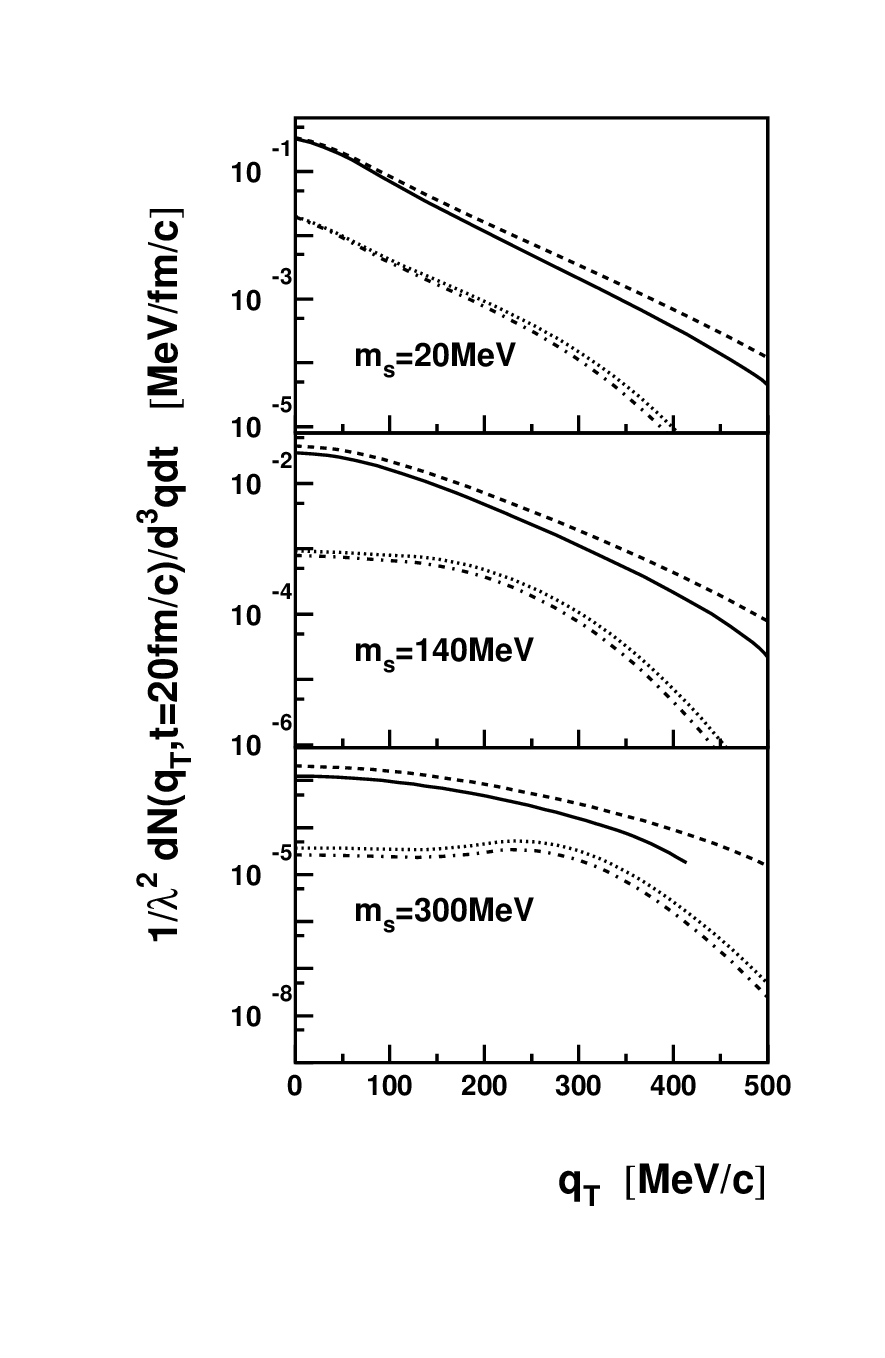,width=0.6\textwidth}
\vspace{0.5cm}
\end{center}
\caption{Rate of the production of mesons as a
function of the transverse momentum at the end of the evolution ($t=20fm/c$).
 The dashed line is the quantum meson production rate for 
 the uncorrelated initial state, the dotted line is the semiclassical
meson production rate for the same final fermion momentum
distribution.
 The solid line is
the quantum meson production rate for the correlated initial state,
the dashed-dotted line is the semiclassical
meson production rate for the same final fermion momentum
distribution. 
  The upper, middle and lower panel 
correspond to the mass of the produced meson of $20$, $140$
and $300MeV$ respectively. The energy of the collision is $180MeV/$nucleon.}
\label{cltr}
\end{figure}

\newpage
\begin{figure}
\begin{center}
\epsfig{file=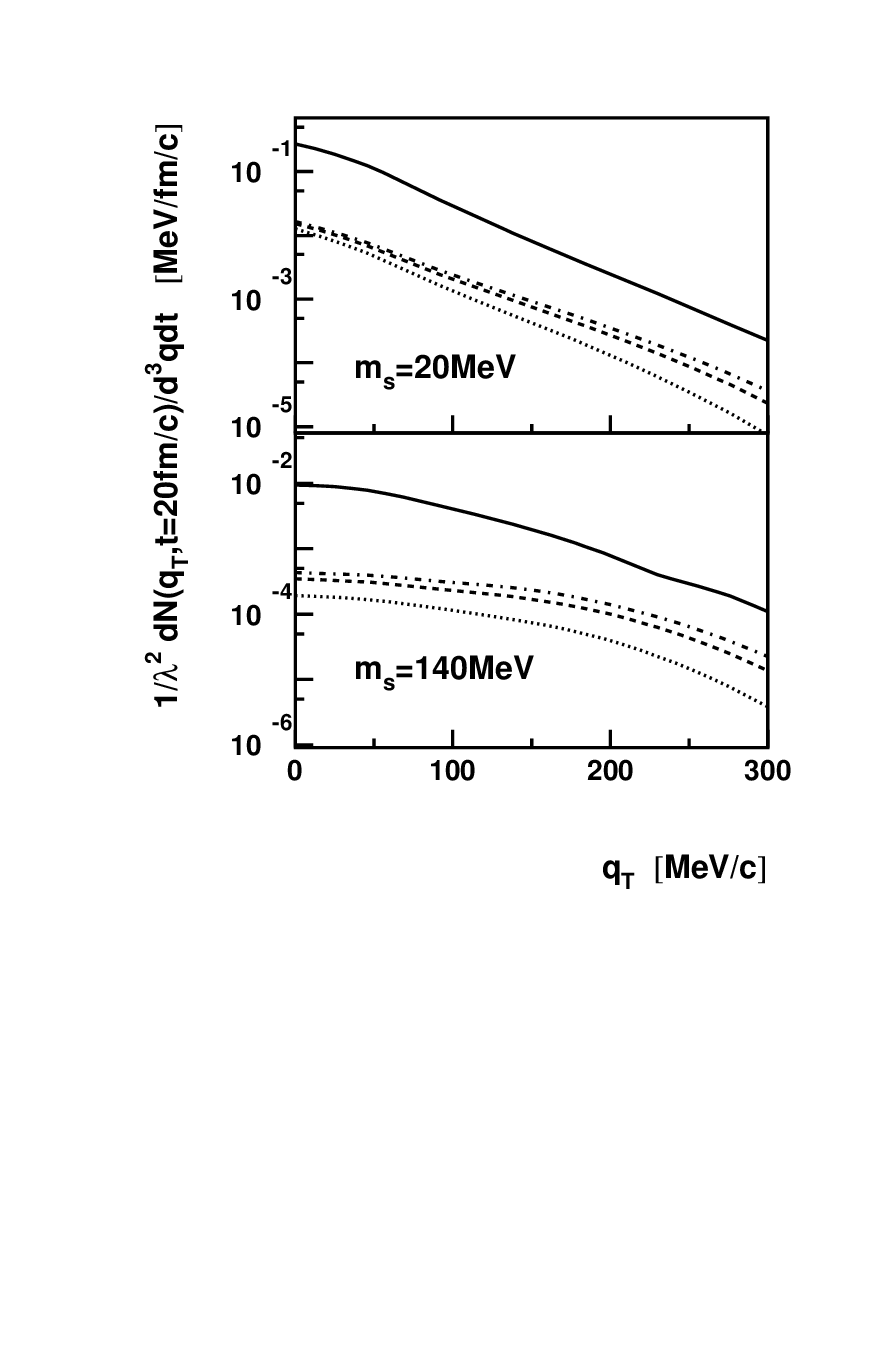,width=0.6\textwidth}
\vspace{0.5cm}
\end{center}
\caption{Rate of the production of mesons as a
function of the transverse momentum at the end of the evolution ($t=20fm/c$).
  The solid line is
the quantum  meson production rate for the correlated initial state,
the dashed-dotted line is the semiclassical
meson production rate for the same final fermion momentum
distribution. 
The dashed line is the semiclassical meson production rate for the BUU 
evolved
correlated initial state. 
 The dotted line is the semiclassical meson production rate for the BUU 
evolved
Hartree initial state. 
  The upper and lower panel 
correspond to the mass of the produced meson of $20$ and $140MeV$
 respectively. 
The energy of the collision is $110MeV/$nucleon.}
\label{tr14}
\end{figure}

\newpage
\begin{figure}
\begin{picture}(10000,25000)
  \thicklines
    \seglength=300 \gaplength=300
  \drawline\scalar[\E\REG](0,19500)[8]
  \addtolength{\scalarbackx}{4000}
  \put(\scalarbackx,\scalarbacky){\oval(8000,4000)} 
  \addtolength{\scalarbacky}{2000}
  \put(\scalarbackx,\scalarbacky){\vector(1,0){0}}
  \addtolength{\scalarbacky}{-4000}
  \put(\scalarbackx,\scalarbacky){\vector(-1,0){0}}
  \addtolength{\scalarbacky}{2000}
  \addtolength{\scalarbackx}{4000}
    \seglength=300 \gaplength=300
    \drawline\scalar[\E\REG](\scalarbackx,\scalarbacky)[8]
  \addtolength{\scalarfronty}{2000}
  \addtolength{\scalarfrontx}{-2000}
    \drawline\photon[\N\FLIPPED](\scalarfrontx,\scalarfronty)[3]
  \addtolength{\scalarfrontx}{-4000}
    \drawline\photon[\N\REG](\scalarfrontx,\scalarfronty)[3]
  \addtolength{\photonbackx}{2000}
    \put(\photonbackx,\photonbacky){\oval(4000,2000)} 
  \addtolength{\photonbacky}{1000}
  \put(\photonbackx,\photonbacky){\vector(1,0){0}}
  \addtolength{\photonbacky}{-2000}
  \put(\photonbackx,\photonbacky){\vector(-1,0){0}}
    \seglength=300 \gaplength=300
  \drawline\scalar[\E\REG](0,6500)[8]
  \addtolength{\scalarbackx}{4000}
  \put(\scalarbackx,\scalarbacky){\oval(8000,4000)} 
  \addtolength{\scalarbacky}{2000}
  \put(\scalarbackx,\scalarbacky){\vector(-1,0){0}}
  \addtolength{\scalarbacky}{-4000}
  \put(\scalarbackx,\scalarbacky){\vector(1,0){0}}
  \addtolength{\scalarbacky}{2000}
  \addtolength{\scalarbackx}{4000}
    \seglength=300 \gaplength=300
    \drawline\scalar[\E\REG](\scalarbackx,\scalarbacky)[8]
  \addtolength{\scalarfronty}{2000}
  \addtolength{\scalarfrontx}{-2000}
    \drawline\photon[\N\FLIPPED](\scalarfrontx,\scalarfronty)[3]
  \addtolength{\scalarfrontx}{-4000}
    \drawline\photon[\N\REG](\scalarfrontx,\scalarfronty)[3]
  \addtolength{\photonbackx}{2000}
    \put(\photonbackx,\photonbacky){\oval(4000,2000)} 
  \addtolength{\photonbacky}{1000}
  \put(\photonbackx,\photonbacky){\vector(1,0){0}}
  \addtolength{\photonbacky}{-2000}
  \put(\photonbackx,\photonbacky){\vector(-1,0){0}}
  \seglength=1416 \gaplength=800
   \addtolength{\photonbacky}{2500}
    \drawline\scalar[\S\REG](\photonbackx,\photonbacky)[5]
   \addtolength{\photonbacky}{13000}
  \seglength=1416 \gaplength=800
    \drawline\scalar[\S\REG](\photonbackx,\photonbacky)[5]
\end{picture}
\caption{The cut diagrams for the meson self-energy consisting of 
self-energy insertions on the fermion lines in the one-loop diagram.
The wiggly line denotes the time local interaction, the cut is
indicated by the long dashed line, other symbols as in Fig. \ref{one_loo}.}
\label{diag_se}
\end{figure}
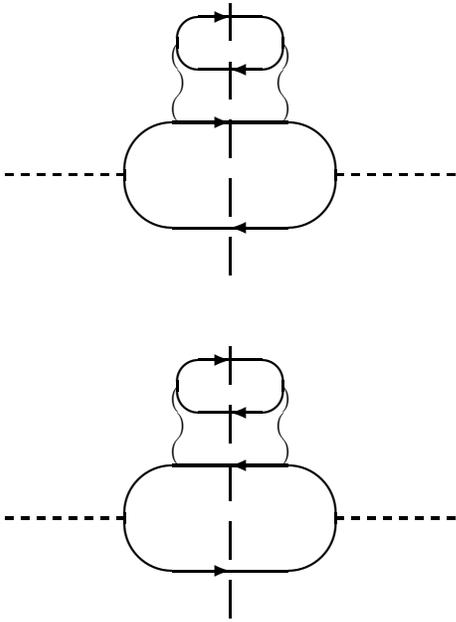

\newpage
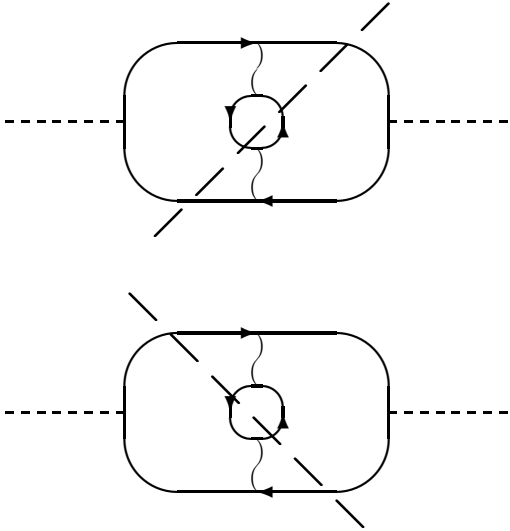
\begin{figure}
\begin{picture}(10000,24000)
  \thicklines
    \seglength=300 \gaplength=300
  \drawline\scalar[\E\REG](0,19000)[8]
  \addtolength{\scalarbackx}{5000}
  \put(\scalarbackx,\scalarbacky){\oval(10000,6000)} 
  \addtolength{\scalarbacky}{3000}
    \drawline\photon[\S\REG](\scalarbackx,\scalarbacky)[2]
  \put(\scalarbackx,\scalarbacky){\vector(1,0){0}}
  \addtolength{\scalarbacky}{-6000}
    \drawline\photon[\N\REG](\scalarbackx,\scalarbacky)[2]
  \put(\scalarbackx,\scalarbacky){\vector(-1,0){0}}
  \addtolength{\scalarbacky}{3000}
  \put(\scalarbackx,\scalarbacky){\oval(2000,2000)} 
  \addtolength{\scalarbackx}{1000}
  \put(\scalarbackx,\scalarbacky){\vector(0,1){0}}
  \addtolength{\scalarbackx}{-2000}
  \put(\scalarbackx,\scalarbacky){\vector(0,-1){0}}
  \addtolength{\scalarbackx}{6000}
    \seglength=300 \gaplength=300
  \drawline\scalar[\E\REG](\scalarbackx,\scalarbacky)[8]
    \seglength=1416 \gaplength=800
  \addtolength{\scalarfronty}{4500}
    \drawline\scalar[\SW\REG](\scalarfrontx,\scalarfronty)[6]
    \seglength=300 \gaplength=300
  \drawline\scalar[\E\REG](0,8000)[8]
  \addtolength{\scalarbackx}{5000}
  \put(\scalarbackx,\scalarbacky){\oval(10000,6000)} 
  \addtolength{\scalarbacky}{3000}
    \drawline\photon[\S\REG](\scalarbackx,\scalarbacky)[2]
  \put(\scalarbackx,\scalarbacky){\vector(1,0){0}}
  \addtolength{\scalarbacky}{-6000}
    \drawline\photon[\N\REG](\scalarbackx,\scalarbacky)[2]
  \put(\scalarbackx,\scalarbacky){\vector(-1,0){0}}
  \addtolength{\scalarbacky}{3000}
  \put(\scalarbackx,\scalarbacky){\oval(2000,2000)} 
  \addtolength{\scalarbackx}{1000}
  \put(\scalarbackx,\scalarbacky){\vector(0,1){0}}
  \addtolength{\scalarbackx}{-2000}
  \put(\scalarbackx,\scalarbacky){\vector(0,-1){0}}
  \addtolength{\scalarbackx}{6000}
    \seglength=300 \gaplength=300
  \drawline\scalar[\E\REG](\scalarbackx,\scalarbacky)[8]
    \seglength=1416 \gaplength=800
  \addtolength{\scalarfronty}{4500}
  \addtolength{\scalarfrontx}{-9800}
    \drawline\scalar[\SE\REG](\scalarfrontx,\scalarfronty)[6]
\end{picture}
\caption{The cut diagrams for the meson self-energy which are not
self-energy insertions on the fermion lines in the one-loop
diagram. Symbols as in Figs. \ref{one_loo} and \ref{diag_se}.}
\label{diag_inne}
\end{figure}

\newpage
\begin{figure}
\begin{picture}(10000,50000)
  \thicklines
   \seglength=300 \gaplength=300
   \drawline\scalar[\E\REG](0,42000)[8]
   \drawline\fermion[\NE\REG](\scalarbackx,\scalarbacky)[5657]
   \drawline\fermion[\SE\REG](\scalarbackx,\scalarbacky)[5657]
  \addtolength{\scalarbackx}{2000}
  \addtolength{\scalarbacky}{2000}
  \put(\scalarbackx,\scalarbacky){\vector(1,1){0}}
  \addtolength{\scalarbacky}{-4000}
  \put(\scalarbackx,\scalarbacky){\vector(-1,1){0}}
   \drawline\fermion[\N\REG](\fermionbackx,\fermionbacky)[8000]
  \addtolength{\scalarbackx}{2000}
  \addtolength{\scalarbacky}{2000}
  \put(\scalarbackx,\scalarbacky){\vector(0,-1){0}}
  \addtolength{\scalarbacky}{4000}
   \drawline\photon[\E\REG](\scalarbackx,\scalarbacky)[4]
  \addtolength{\scalarbacky}{-8000}
   \drawline\photon[\E\REG](\scalarbackx,\scalarbacky)[4]
   \drawline\fermion[\N\REG](\photonbackx,\photonbacky)[8000]
  \addtolength{\photonbacky}{4000}
  \put(\photonbackx,\photonbacky){\vector(0,-1){0}}
  \addtolength{\photonbacky}{4000}
   \drawline\fermion[\SE\REG](\fermionbackx,\fermionbacky)[5657]
   \drawline\fermion[\SW\REG](\fermionbackx,\fermionbacky)[5657]
  \addtolength{\photonbacky}{-2000}
  \addtolength{\photonbackx}{2000}
  \put(\photonbackx,\photonbacky){\vector(-1,1){0}}
  \addtolength{\photonbacky}{-4000}
  \put(\photonbackx,\photonbacky){\vector(1,1){0}}
  \addtolength{\photonbacky}{2000}
  \addtolength{\photonbackx}{2000}
     \seglength=300 \gaplength=300
     \drawline\scalar[\E\REG](\photonbackx,\photonbacky)[8]
  \addtolength{\photonbacky}{5200}
  \addtolength{\photonbackx}{-800}
     \seglength=1416 \gaplength=800
     \drawline\scalar[\SW\REG](\photonbackx,\photonbacky)[6]
   \seglength=300 \gaplength=300
   \drawline\scalar[\E\REG](0,30000)[8]
   \drawline\fermion[\NE\REG](\scalarbackx,\scalarbacky)[5657]
   \drawline\fermion[\SE\REG](\scalarbackx,\scalarbacky)[5657]
  \addtolength{\scalarbackx}{2000}
  \addtolength{\scalarbacky}{2000}
  \put(\scalarbackx,\scalarbacky){\vector(1,1){0}}
  \addtolength{\scalarbacky}{-4000}
  \put(\scalarbackx,\scalarbacky){\vector(-1,1){0}}
   \drawline\fermion[\N\REG](\fermionbackx,\fermionbacky)[8000]
  \addtolength{\scalarbackx}{2000}
  \addtolength{\scalarbacky}{2000}
  \put(\scalarbackx,\scalarbacky){\vector(0,-1){0}}
  \addtolength{\scalarbacky}{4000}
   \drawline\photon[\E\REG](\scalarbackx,\scalarbacky)[4]
  \addtolength{\scalarbacky}{-8000}
   \drawline\photon[\E\REG](\scalarbackx,\scalarbacky)[4]
   \drawline\fermion[\N\REG](\photonbackx,\photonbacky)[8000]
  \addtolength{\photonbacky}{4000}
  \put(\photonbackx,\photonbacky){\vector(0,-1){0}}
  \addtolength{\photonbacky}{4000}
   \drawline\fermion[\SE\REG](\fermionbackx,\fermionbacky)[5657]
   \drawline\fermion[\SW\REG](\fermionbackx,\fermionbacky)[5657]
  \addtolength{\photonbacky}{-2000}
  \addtolength{\photonbackx}{2000}
  \put(\photonbackx,\photonbacky){\vector(-1,1){0}}
  \addtolength{\photonbacky}{-4000}
  \put(\photonbackx,\photonbacky){\vector(1,1){0}}
  \addtolength{\photonbacky}{2000}
  \addtolength{\photonbackx}{2000}
     \seglength=300 \gaplength=300
     \drawline\scalar[\E\REG](\photonbackx,\photonbacky)[8]
  \addtolength{\photonfronty}{8500}
  \addtolength{\photonfrontx}{-2500}
     \seglength=1416 \gaplength=800
     \drawline\scalar[\SE\REG](\photonfrontx,\photonfronty)[6]
   \seglength=300 \gaplength=300
   \drawline\scalar[\E\REG](0,19000)[8]
   \drawline\fermion[\NE\REG](\scalarbackx,\scalarbacky)[5657]
   \drawline\fermion[\SE\REG](\scalarbackx,\scalarbacky)[5657]
  \addtolength{\scalarbackx}{2000}
  \addtolength{\scalarbacky}{2000}
  \put(\scalarbackx,\scalarbacky){\vector(-1,-1){0}}
  \addtolength{\scalarbacky}{-4000}
  \put(\scalarbackx,\scalarbacky){\vector(1,-1){0}}
   \drawline\fermion[\N\REG](\fermionbackx,\fermionbacky)[8000]
  \addtolength{\scalarbackx}{2000}
  \addtolength{\scalarbacky}{2000}
  \put(\scalarbackx,\scalarbacky){\vector(0,1){0}}
  \addtolength{\scalarbacky}{4000}
   \drawline\photon[\E\REG](\scalarbackx,\scalarbacky)[4]
  \addtolength{\scalarbacky}{-8000}
   \drawline\photon[\E\REG](\scalarbackx,\scalarbacky)[4]
   \drawline\fermion[\N\REG](\photonbackx,\photonbacky)[8000]
  \addtolength{\photonbacky}{4000}
  \put(\photonbackx,\photonbacky){\vector(0,-1){0}}
  \addtolength{\photonbacky}{4000}
   \drawline\fermion[\SE\REG](\fermionbackx,\fermionbacky)[5657]
   \drawline\fermion[\SW\REG](\fermionbackx,\fermionbacky)[5657]
  \addtolength{\photonbacky}{-2000}
  \addtolength{\photonbackx}{2000}
  \put(\photonbackx,\photonbacky){\vector(-1,1){0}}
  \addtolength{\photonbacky}{-4000}
  \put(\photonbackx,\photonbacky){\vector(1,1){0}}
  \addtolength{\photonbacky}{2000}
  \addtolength{\photonbackx}{2000}
     \seglength=300 \gaplength=300
     \drawline\scalar[\E\REG](\photonbackx,\photonbacky)[8]
  \addtolength{\photonbacky}{5200}
  \addtolength{\photonbackx}{-800}
     \seglength=1416 \gaplength=800
     \drawline\scalar[\SW\REG](\photonbackx,\photonbacky)[6]
   \seglength=300 \gaplength=300
   \drawline\scalar[\E\REG](0,9000)[8]
   \drawline\fermion[\NE\REG](\scalarbackx,\scalarbacky)[5657]
   \drawline\fermion[\SE\REG](\scalarbackx,\scalarbacky)[5657]
  \addtolength{\scalarbackx}{2000}
  \addtolength{\scalarbacky}{2000}
  \put(\scalarbackx,\scalarbacky){\vector(-1,-1){0}}
  \addtolength{\scalarbacky}{-4000}
  \put(\scalarbackx,\scalarbacky){\vector(1,-1){0}}
   \drawline\fermion[\N\REG](\fermionbackx,\fermionbacky)[8000]
  \addtolength{\scalarbackx}{2000}
  \addtolength{\scalarbacky}{2000}
  \put(\scalarbackx,\scalarbacky){\vector(0,1){0}}
  \addtolength{\scalarbacky}{4000}
   \drawline\photon[\E\REG](\scalarbackx,\scalarbacky)[4]
  \addtolength{\scalarbacky}{-8000}
   \drawline\photon[\E\REG](\scalarbackx,\scalarbacky)[4]
   \drawline\fermion[\N\REG](\photonbackx,\photonbacky)[8000]
  \addtolength{\photonbacky}{4000}
  \put(\photonbackx,\photonbacky){\vector(0,-1){0}}
  \addtolength{\photonbacky}{4000}
   \drawline\fermion[\SE\REG](\fermionbackx,\fermionbacky)[5657]
   \drawline\fermion[\SW\REG](\fermionbackx,\fermionbacky)[5657]
  \addtolength{\photonbacky}{-2000}
  \addtolength{\photonbackx}{2000}
  \put(\photonbackx,\photonbacky){\vector(-1,1){0}}
  \addtolength{\photonbacky}{-4000}
  \put(\photonbackx,\photonbacky){\vector(1,1){0}}
  \addtolength{\photonbacky}{2000}
  \addtolength{\photonbackx}{2000}
     \seglength=300 \gaplength=300
     \drawline\scalar[\E\REG](\photonbackx,\photonbacky)[8]
  \addtolength{\photonfronty}{8500}
  \addtolength{\photonfrontx}{-2500}
     \seglength=1416 \gaplength=800
     \drawline\scalar[\SE\REG](\photonfrontx,\photonfronty)[6]
\end{picture}
\caption{The cut diagrams for the meson self-energy which are not
self-energy insertions on the fermion lines in the one-loop diagram.
These diagrams are zero for the pion-nucleon interaction vertex.
Symbols as in Figs. \ref{one_loo} and \ref{diag_se}.}
\label{diag_ze}
\end{figure}
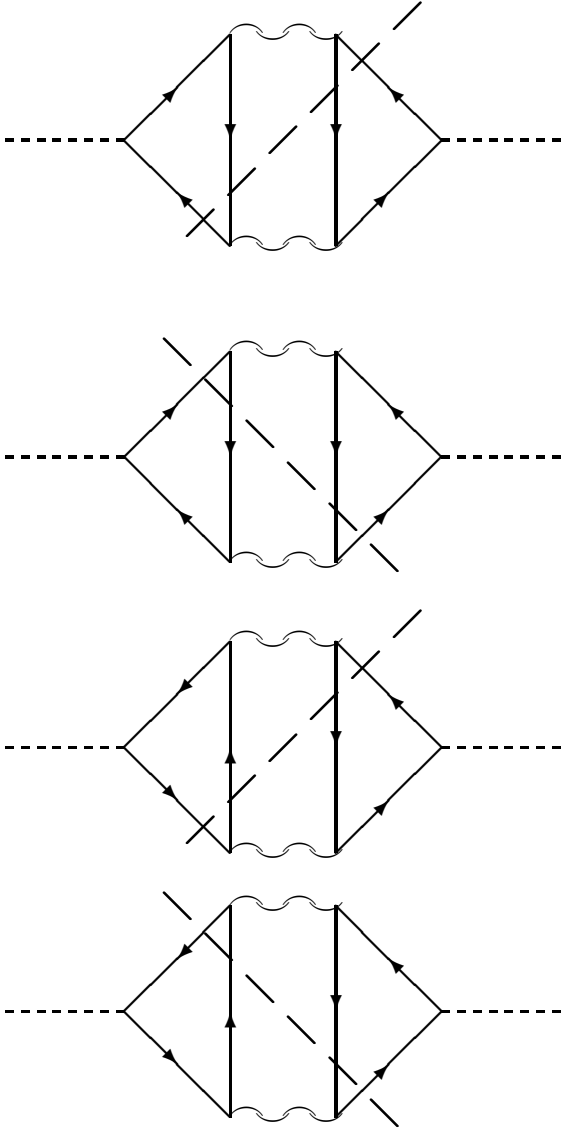

\newpage
\begin{figure}
\begin{picture}(15000,25000)
  \thicklines
    \seglength=300 \gaplength=300
  \drawline\scalar[\E\REG](0,10500)[8]
  \addtolength{\scalarbackx}{8000}
  \put(\scalarbackx,\scalarbacky){\oval(16000,8000)} 
  \addtolength{\scalarbacky}{4000}
  \put(\scalarbackx,\scalarbacky){\vector(1,0){0}}
  \addtolength{\scalarbacky}{-8000}
  \put(\scalarbackx,\scalarbacky){\vector(-1,0){0}}
  \addtolength{\scalarbacky}{4000}
  \addtolength{\scalarbackx}{8000}
    \seglength=300 \gaplength=300
    \drawline\scalar[\E\REG](\scalarbackx,\scalarbacky)[8]
  \addtolength{\scalarfronty}{4000}
  \addtolength{\scalarfrontx}{-4000}
    \drawline\photon[\N\FLIPPED](\scalarfrontx,\scalarfronty)[3]
  \addtolength{\scalarfrontx}{-8000}
    \drawline\photon[\N\REG](\scalarfrontx,\scalarfronty)[3]
  \addtolength{\photonbackx}{4000}
    \put(\photonbackx,\photonbacky){\oval(8000,4000)} 
  \addtolength{\photonbacky}{2000}
  \put(\photonbackx,\photonbacky){\vector(1,0){0}}
  \addtolength{\photonbacky}{-4000}
  \put(\photonbackx,\photonbacky){\vector(-1,0){0}}
   \addtolength{\photonbacky}{4000}
  \addtolength{\photonbackx}{-2000}
    \drawline\photon[\N\REG](\photonbackx,\photonbacky)[3]
  \addtolength{\photonfrontx}{4000}
    \drawline\photon[\N\FLIPPED](\photonfrontx,\photonfronty)[3]
  \addtolength{\photonbackx}{-2000}
    \put(\photonbackx,\photonbacky){\oval(4000,2000)} 
  \addtolength{\photonbacky}{1000}
  \put(\photonbackx,\photonbacky){\vector(1,0){0}}
  \addtolength{\photonbacky}{-2000}
  \put(\photonbackx,\photonbacky){\vector(-1,0){0}}
   \addtolength{\photonbacky}{3000}
  \seglength=1416 \gaplength=800
    \drawline\scalar[\S\REG](\photonbackx,\photonbacky)[9]
\end{picture}
\caption{A cut diagram for the meson self-energy of a type 
of direct Born self energy insertions on fermion lines. Many other 
self energy insertion can be added, giving after cutting
 processes involving many on-shell fermions.
Symbols as in Figs. \ref{one_loo} and \ref{diag_se}.}
%\label{se-0Ö ¥Ã}$)B
\end{figure}
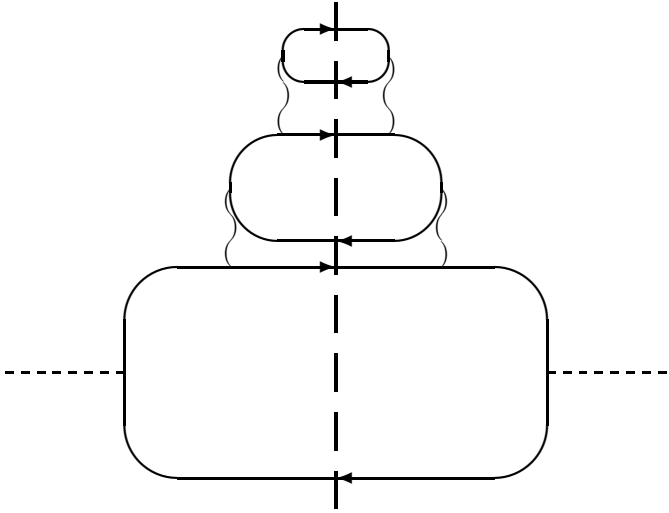

\newpage
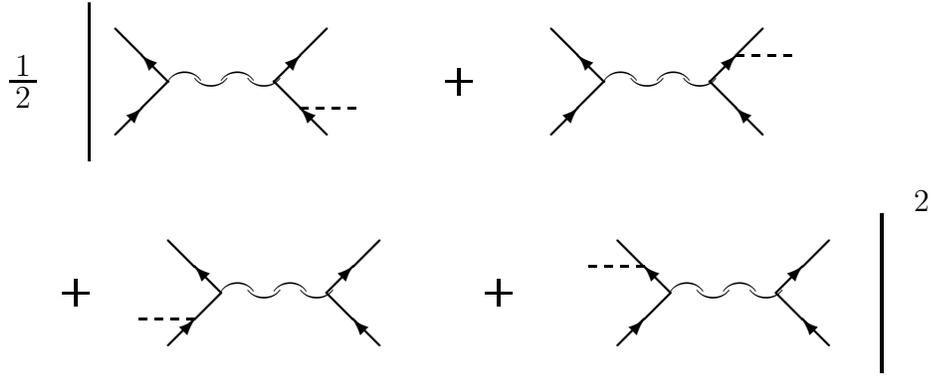
\begin{figure}
  \begin{picture}(10000,22000)
  \thicklines
   \drawline\photon[\E\REG](4000,15000)[4]
   \drawline\fermion[\NE\REG](\photonbackx,\photonbacky)[2828]
  \put(\particlemidx,\particlemidy){\vector(1,1){0}}
   \drawline\fermion[\SE\REG](\photonbackx,\photonbacky)[2828]
  \put(\particlemidx,\particlemidy){\vector(-1,1){0}}
    \seglength=300 \gaplength=300
  \drawline\scalar[\E\REG](\particlemidx,\particlemidy)[4]
   \drawline\fermion[\NW\REG](\photonfrontx,\photonfronty)[2828]
  \put(\particlemidx,\particlemidy){\vector(-1,1){0}}
   \drawline\fermion[\SW\REG](\photonfrontx,\photonfronty)[2828]
  \put(\particlemidx,\particlemidy){\vector(1,1){0}}
  \addtolength{\photonfrontx}{-3000}
  \addtolength{\photonfronty}{-3000}
   \drawline\fermion[\N\REG](\photonfrontx,\photonfronty)[6000]
  \addtolength{\photonfrontx}{-3000}
  \addtolength{\photonfronty}{3000}
   \drawline\fermion[\E\REG](\photonfrontx,\photonfronty)[1000]
  \addtolength{\photonfrontx}{-800}
  \addtolength{\photonfronty}{200}
    \put(\photonfrontx,\photonfronty){\makebox(1000,1000)\Huge{1}}
  \addtolength{\photonfronty}{-1200}
  \put(\photonfrontx,\photonfronty){\makebox(1000,1000)\Huge{2}}
  \addtolength{\photonbackx}{6500}
   \drawline\fermion[\E\REG](\photonbackx,\photonbacky)[1000]
  \addtolength{\photonbackx}{500}
  \addtolength{\photonbacky}{-500}
   \drawline\fermion[\N\REG](\photonbackx,\photonbacky)[1000]
  \addtolength{\photonbackx}{5500}
  \addtolength{\photonbacky}{500}
   \drawline\photon[\E\REG](\photonbackx,\photonbacky)[4]
   \drawline\fermion[\NE\REG](\photonbackx,\photonbacky)[2828]
  \put(\particlemidx,\particlemidy){\vector(1,1){0}}
    \seglength=300 \gaplength=300
  \drawline\scalar[\E\REG](\particlemidx,\particlemidy)[4]
   \drawline\fermion[\SE\REG](\photonbackx,\photonbacky)[2828]
  \put(\particlemidx,\particlemidy){\vector(-1,1){0}}
   \drawline\fermion[\NW\REG](\photonfrontx,\photonfronty)[2828]
  \put(\particlemidx,\particlemidy){\vector(-1,1){0}}
   \drawline\fermion[\SW\REG](\photonfrontx,\photonfronty)[2828]
  \put(\particlemidx,\particlemidy){\vector(1,1){0}}
   \drawline\photon[\E\REG](6000,7000)[4]
   \drawline\fermion[\NE\REG](\photonbackx,\photonbacky)[2828]
  \put(\particlemidx,\particlemidy){\vector(1,1){0}}
   \drawline\fermion[\SE\REG](\photonbackx,\photonbacky)[2828]
  \put(\particlemidx,\particlemidy){\vector(-1,1){0}}
   \drawline\fermion[\NW\REG](\photonfrontx,\photonfronty)[2828]
  \put(\particlemidx,\particlemidy){\vector(-1,1){0}}
    \drawline\fermion[\SW\REG](\photonfrontx,\photonfronty)[2828]
  \put(\particlemidx,\particlemidy){\vector(1,1){0}}
    \seglength=300 \gaplength=300
  \drawline\scalar[\W\REG](\particlemidx,\particlemidy)[4]
  \addtolength{\photonbackx}{4000}
  \addtolength{\photonfrontx}{-5500}
  \addtolength{\photonfronty}{-500}
   \drawline\fermion[\N\REG](\photonfrontx,\photonfronty)[1000]
  \addtolength{\photonfrontx}{-500}
  \addtolength{\photonfronty}{500}
   \drawline\fermion[\E\REG](\photonfrontx,\photonfronty)[1000]
  \addtolength{\photonbackx}{1000}
  \addtolength{\photonbackx}{1000}
   \drawline\fermion[\E\REG](\photonbackx,\photonbacky)[1000]
  \addtolength{\photonbackx}{500}
  \addtolength{\photonbacky}{-500}
   \drawline\fermion[\N\REG](\photonbackx,\photonbacky)[1000]
  \addtolength{\photonbackx}{6500}
  \addtolength{\photonbacky}{500}
   \drawline\photon[\E\REG](\photonbackx,\photonbacky)[4]
   \drawline\fermion[\NE\REG](\photonbackx,\photonbacky)[2828]
  \put(\particlemidx,\particlemidy){\vector(1,1){0}}
   \drawline\fermion[\SE\REG](\photonbackx,\photonbacky)[2828]
  \put(\particlemidx,\particlemidy){\vector(-1,1){0}}
   \drawline\fermion[\NW\REG](\photonfrontx,\photonfronty)[2828]
  \put(\particlemidx,\particlemidy){\vector(-1,1){0}}
    \seglength=300 \gaplength=300
  \drawline\scalar[\W\REG](\particlemidx,\particlemidy)[4]
   \drawline\fermion[\SW\REG](\photonfrontx,\photonfronty)[2828]
  \put(\particlemidx,\particlemidy){\vector(1,1){0}}
  \addtolength{\photonbackx}{4000}
  \addtolength{\photonbacky}{-3000}
   \drawline\fermion[\N\REG](\photonbackx,\photonbacky)[6000]
  \addtolength{\photonbackx}{1000}
  \addtolength{\photonbacky}{6000}
  \put(\photonbackx,\photonbacky){\makebox(1000,1000){2}}
\end{picture}
\caption{The matrix element from the cut second order direct
diagrams. See text for details.}
\label{mat_fu}
\end{figure}

\newpage
\begin{figure}
  \begin{picture}(10000,20000)
  \thicklines
   \drawline\photon[\E\REG](2000,15000)[4]
   \drawline\fermion[\NE\REG](\photonbackx,\photonbacky)[2828]
  \put(\particlemidx,\particlemidy){\vector(1,1){0}}
   \drawline\fermion[\SE\REG](\photonbackx,\photonbacky)[2828]
  \put(\particlemidx,\particlemidy){\vector(-1,1){0}}
    \seglength=300 \gaplength=300
  \drawline\scalar[\E\REG](\particlemidx,\particlemidy)[4]
   \drawline\fermion[\NW\REG](\photonfrontx,\photonfronty)[2828]
  \put(\particlemidx,\particlemidy){\vector(-1,1){0}}
   \drawline\fermion[\SW\REG](\photonfrontx,\photonfronty)[2828]
  \put(\particlemidx,\particlemidy){\vector(1,1){0}}
  \addtolength{\photonbackx}{4000}
  \addtolength{\photonbacky}{-3000}
   \drawline\fermion[\N\REG](\photonbackx,\photonbacky)[6000]
  \addtolength{\photonfrontx}{-3000}
  \addtolength{\photonfronty}{-3000}
   \drawline\fermion[\N\REG](\photonfrontx,\photonfronty)[6000]
  \addtolength{\photonbackx}{1000}
  \addtolength{\photonbacky}{6000}
  \put(\photonbackx,\photonbacky){\makebox(1000,1000){2}}
  \addtolength{\photonbackx}{1000}
  \addtolength{\photonbacky}{-3000}
   \drawline\fermion[\E\REG](\photonbackx,\photonbacky)[1000]
  \addtolength{\photonbackx}{500}
  \addtolength{\photonbacky}{-500}
   \drawline\fermion[\N\REG](\photonbackx,\photonbacky)[1000]
  \addtolength{\photonbackx}{5500}
  \addtolength{\photonbacky}{500}
   \drawline\photon[\E\REG](\photonbackx,\photonbacky)[4]
   \drawline\fermion[\NE\REG](\photonbackx,\photonbacky)[2828]
  \put(\particlemidx,\particlemidy){\vector(1,1){0}}
    \seglength=300 \gaplength=300
  \drawline\scalar[\E\REG](\particlemidx,\particlemidy)[4]
   \drawline\fermion[\SE\REG](\photonbackx,\photonbacky)[2828]
  \put(\particlemidx,\particlemidy){\vector(-1,1){0}}
   \drawline\fermion[\NW\REG](\photonfrontx,\photonfronty)[2828]
  \put(\particlemidx,\particlemidy){\vector(-1,1){0}}
   \drawline\fermion[\SW\REG](\photonfrontx,\photonfronty)[2828]
  \put(\particlemidx,\particlemidy){\vector(1,1){0}}
  \addtolength{\photonbackx}{4000}
  \addtolength{\photonbacky}{-3000}
   \drawline\fermion[\N\REG](\photonbackx,\photonbacky)[6000]
  \addtolength{\photonfrontx}{-3000}
  \addtolength{\photonfronty}{-3000}
   \drawline\fermion[\N\REG](\photonfrontx,\photonfronty)[6000]
  \addtolength{\photonbackx}{1000}
  \addtolength{\photonbacky}{6000}
  \put(\photonbackx,\photonbacky){\makebox(1000,1000){2}}
\end{picture}
\caption{The matrix element from the cut  fermion self-energy insertion
diagrams
from Fig \ref{diag_se}. See text for details.}
\label{mat_se}
\end{figure}
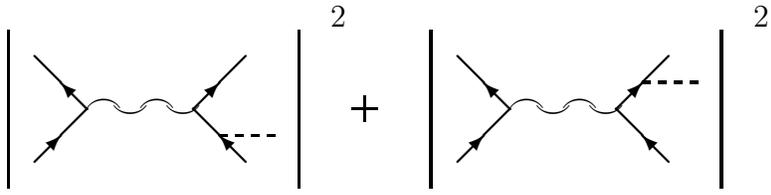

\newpage
\begin{figure}
\begin{center}
\epsfig{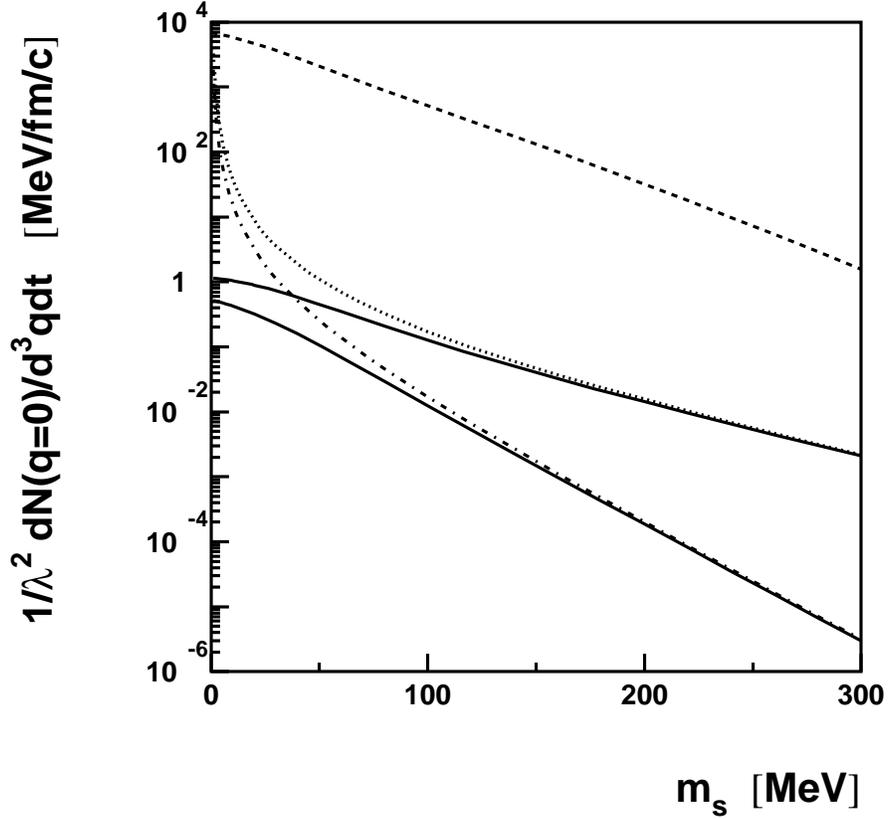}
\vspace{0.5cm}
\end{center}
\caption{Rate of meson production at zero momentum
 as a function of the meson mass. The dashed line is the quantum
one-loop result (\ref{ol_eq}). 
The dotted line is the semiclassical production rate
with the quantum equilibrium fermion momentum distribution
and the dashed-dotted line is the the semiclassical production rate
with the semiclassical equilibrium fermion momentum distribution ($\gamma
\rightarrow 0$). 
The solid line close to the dotted line is the same
as the dotted line but with a non-zero width 
($\gamma=60MeV$) of the retarded propagator in the
calculation of the semiclassical production cross section. 
The solid line close to the dashed-dotted line is the same
as the dashed-dotted
 line but with a non-zero width ($\gamma=60MeV$) of the 
retarded propagator in the
calculation of the semiclassical production cross section. 
The Fermi energy, the temperature and the width are $40$, $30$ and $60MeV$
respectively.}
\label{proste_fig}
\end{figure}


\begin{references}
\bibitem{com} The correlated initial state is not the stationary state
of the BUU evolution equation. Thus, its use as the initial state is
not consistent with the time dependent part of the evolution governed
the BUU collision integral. However, it is motivated by the known
quantum effects in the ground state fermion momentum distribution in nuclei.
\end{references}
\end{document}